%% file: main.tex
\documentclass[10pt,journal,compsoc]{IEEEtran}

\input{packages}
\input{macros}

\usepackage{url}

\begin{document}

\title{NCQ: Code reuse support for node.js developers}
\author{Brittany Reid, Marcelo d'Amorim, Markus Wagner and Christoph Treude
\IEEEcompsocitemizethanks{%
\IEEEcompsocthanksitem Brittany Reid and Markus Wagner are with the University of Adelaide, Australia.\protect\\
E-mail: 
\{brittany.reid, markus.wagner\}@adelaide.edu.au%
\IEEEcompsocthanksitem Marcelo d'Amorim is with the North Carolina State University, USA, and with the Universidade Federal de Pernambuco, Brazil\protect\\
E-mail: mdamori@ncsu.edu%
\IEEEcompsocthanksitem Christoph Treude is with the University of Melbourne, Australia\protect\\
E-mail: christoph.treude@unimelb.edu.au%
}%
}

\markboth{Submitted to IEEE Transactions on Software Engineering}%
{Shell \MakeLowercase{\textit{et al.}}: Bare Demo of IEEEtran.cls for Computer Society Journals}

\IEEEtitleabstractindextext{
\begin{abstract}
Code reuse is an important part of software development. The adoption of code reuse practices is especially common among Node.js developers. The Node.js package manager, NPM, indexes over 1 Million packages and developers often seek out packages to solve programming tasks. Due to the vast number of packages, selecting the right package is difficult and time consuming. With the goal of improving productivity of developers that heavily reuse code through third-party packages, we present \textit{Node Code Query} (\tname{}), a \hl{Read-Eval-Print-Loop} environment that allows developers to 1)~search for NPM packages using natural language queries, 2)~search for code snippets related to those packages, 3)~automatically correct errors in these code snippets, 4)~quickly setup new environments for testing those snippets, and 5)~transition between search and editing modes. In two user studies with a total of 20 participants, we find that participants begin programming faster and conclude tasks faster with \tname{} than with baseline approaches, and that they like, among other features, the search for code snippets and packages. Our results suggest that \tname{} makes Node.js developers more efficient in reusing code.

\end{abstract}

}

\maketitle
\IEEEdisplaynontitleabstractindextext
\IEEEpeerreviewmaketitle

\IEEEraisesectionheading{
    \section{Introduction}\label{sec:introduction}
}

\IEEEPARstart{N}{ode.js} is a popular \javascript{} runtime~\cite{nodejs_stats} often used to develop server-side applications. The Node.js package manager, NPM,\footnote{\url{https://www.npmjs.com/}} hosts over 1 million packages~\cite{snyk_io_blog}, with a typical package recursively depending on several others; on average each package directly depends on 5.9 other packages~\cite{kula2017impact}. Despite such a rich package ecosystem, finding the right package and figuring out how to use it can be time-consuming. The vast number of options means that developers are unable to gain detailed understanding of all potential selections and thus make decisions based on limited data. The typical package search process involves developers searching online (\eg{} using a general purpose search engine or the NPM website), then evaluating relevant packages by~1)~installing the selected package and~2)~testing the functionality of that package (\ie{}, running code snippets) to decide if the package is fit for purpose. If a developer needs to try multiple packages (for example, due to a lack of features)\Ignore{to outdated packages or }, this process can become time intensive.

We conducted a preliminary study to understand the challenges of developers in finding packages. Our survey of 55 Node.js developers revealed that package use is indeed a common part of Node.js development, yet most developers encountered challenges in finding packages, with the most common challenge being insufficient documentation and code examples. To sum up, developers have challenges determining \emph{how to use} the packages they find.

To address the challenges of package search in the Node.js ecosystem, we propose Node Code Query (\tname{}), a custom 
\hl{
Read-Eval-Print-Loop 
}
(REPL)~\cite{sandewall1978programming} which integrates 1)~NPM package search using natural language queries, 2)~code snippet search, 3)~automatic code snippet error correction, 4)~automated setup of environments for testing those snippets, and 5)~the ability to transition between search and editing modes. We conjecture that \tname{} is a useful development aid in an environment where much of coding is related to connecting snippets of code as opposed to developing code from scratch. To reduce the effect of context switching on performance, \tname{} integrates all of the above steps of code reuse into one tool; package and code search, a code editor and a shell to execute code and install packages.

\hl{Existing ideas from the literature in code and library search~}\cite{Campbell2017NLP2CodeCS, ponzanelli2014prompter, zhang2016bing, brandt2010example, ponzanelli2013seahawk, 10.1145/3368089.3417922, NLP2TestableCode}\hl{ inspired \tname{}. The novelty of \tname{} is grounded on the application of these ideas in a different context. We conjecture that \tname{}'s interactive REPL can better assist developers in the common scenarios where they wish to try out new software libraries quickly in an isolated environment. Developers can avoid making changes in their local programming environment when they want to borrow functionality from existing libraries. The combination of REPL, package and code search and code correction is non-trivial; they have been adapted for this new context to be intuitive within a command-line interface, and to reduce friction and context-switching. Our user studies aim to validate these assumptions.}

To validate the usefulness of \tname{}, we evaluated the tool with Node.js developers. Participants' activities were compared against a baseline comprising of internet access and Visual Studio Code, which we identified as the most popular editor for Node.js developers. We asked participants to complete two basic programming tasks, first using the baseline, then using only the tool. Our evaluation shows that participants were able to complete all tasks using \tname{} without context switching, at least as quickly as when they used a well-established editor and had access to online search. Participant feedback for the tool was generally positive in regards to helpfulness, confidence in solutions and tool features. These results suggest that \tname{} is a promising direction to fill an existing gap in tool support for Node.js code reuse, reducing the burden of context switching between searching and editing tools.


\hl{
This paper makes the following contributions:
}

\begin{itemize}[topsep=.2ex,itemsep=.2ex,leftmargin=0.8em]

\item[\Contrib{}]\textbf{Approach.}~A command line tool, \tname{} that integrates documentation from the NPM database, within a REPL with the ability to create temporary environments, search for packages, install those packages, search for and run related code snippets and edit submitted code like a traditional code editor, along with an evaluation in two user studies with a total of 20 participants.


\item[\Contrib{}]\textbf{Artifacts.}~\tname{} is publicly available on GitHub at \url{https://github.com/Brittany-Reid/node_code_query}.

\end{itemize}

\section{Illustrative Example}
\label{sec:example}

Let us consider a scenario where a developer would like to create a CSV file in Node.js. The developer would like to use a package to solve this task; this can be a simple task to manually code, but a package would save time and avoid mistakes. The following sections describe the workflow two developers followed when solving Task 3 of our user study. In the following, we describe the problem-solving process followed by two participants. One participant was asked to use the internet to find packages and to code in Visual Studio Code, while the other participant used only \tnameii{}. Section~\ref{sec:tasks} describes this task in more detail.

\subsection{Typical problem solving in Node.js}
\label{sec:manualexample}


This section describes the workflow of one participant using the internet and a traditional editor. We consider this workflow to be a good example of typical problem solving in Node.js.

\begin{figure}[h!]
    \centering
    \includegraphics[width=0.95\linewidth]{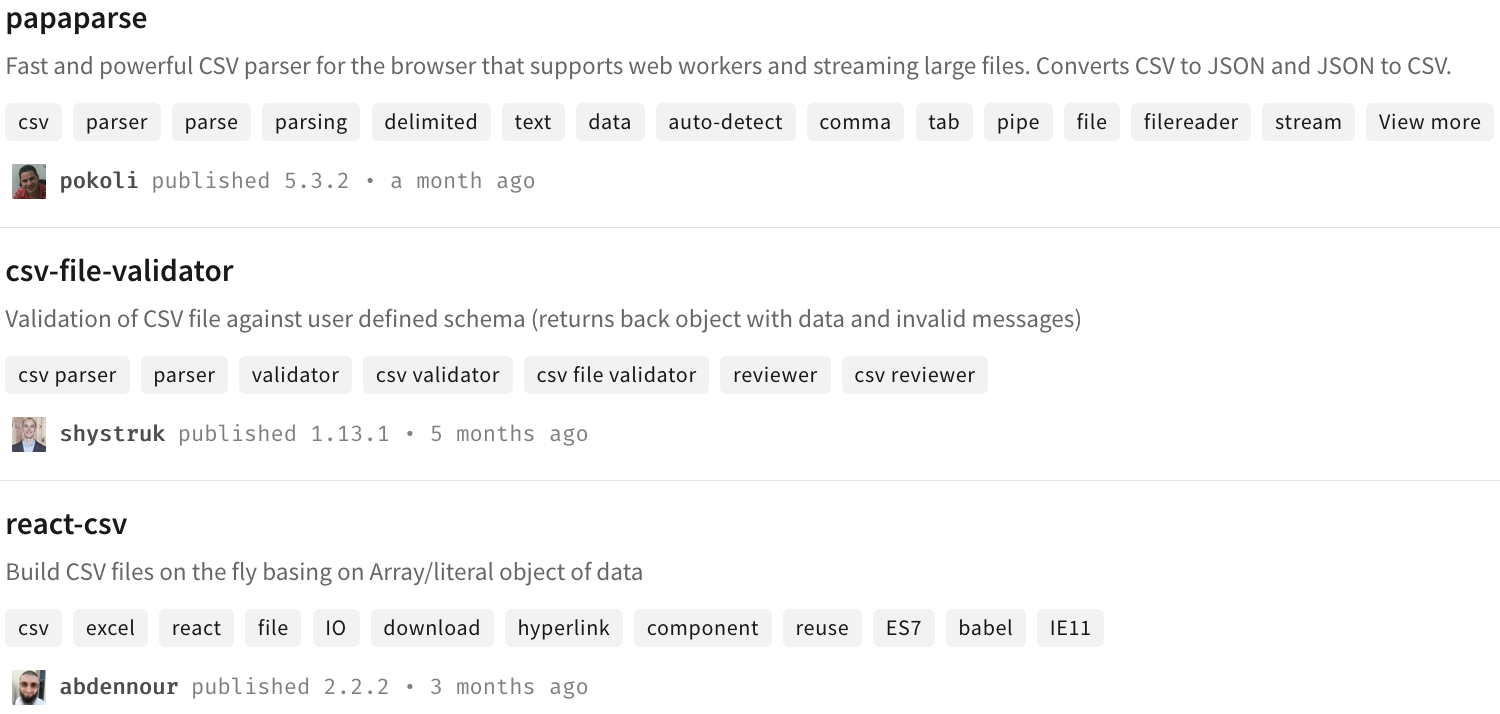}
    \caption{NPM website search results for ``csv file''.}
    \label{fig:npmsearch}
\end{figure}

The developer starts by using a general purpose search engine (all participants in our user study started with a general purpose search engine). In this case, they enter the query ``!npm csv file'' into search engine DuckDuckGo, which redirects them immediately to NPM's search for the query ``csv file''. Figure~\ref{fig:npmsearch} shows the first three results for this query, however, none pertain to \emph{creating} CSV files. After reading the names and descriptions of these packages, the developer refines their search, using the query ``csv file create''. The developer looks through the results for this query, finding the package \texttt{node-create-csv} which mentions converting an array into a CSV file, and selects this package.

After being taken to the README file for this package, the developer can see that it has some usage examples, including the example shown in Figure~\ref{fig:ncc_example} which writes some data to a file.

\begin{figure}[h]
    \begin{lstlisting}[language=diff]
const ObjectsToCsv = require('objects-to-csv');
const data = [{code: 'CA', name: 'California'},
  {code: 'TX', name: 'Texas'},
  {code: 'NY', name: 'New York'},];
(async () => {const csv = new ObjectsToCsv(data);
  await csv.toDisk('./test.csv');
  console.log(await csv.toString());})();
    \end{lstlisting}
    \caption{Usage example from the README for \texttt{node-create-csv}. Note the incorrect package name.}
    \label{fig:ncc_example}
\end{figure}

The developer decides to try to use this package, so they create a new Node.js project. First, they make a new directory for the project, then in that directory, they use the command \texttt{npm init -y} to initialise the project, which creates a special \CodeIn{package.json} file that stores metadata including a list of dependencies. With the project configured, the developer now installs the package with the \texttt{npm install node-create-csv} command.

With the package installed, the developer copies and pastes the example into their editor. 
\hl{They modify the data being written to match the task's requirements,} 
however, when they attempt to run the code, Node.js reports the \texttt{MODULE\_NOT\_FOUND} error despite the package being installed. After some investigation, the developer realises that the example imports a different package, \texttt{objects-to-csv}, and the error refers to \emph{this} package not being installed. \hl{The developer looks over the rest of the examples and sees the same issue, so considers this package to be buggy and moves on.}

\hl{However, during our post-session investigation, we found that \texttt{node-create-csv} is a copy of an existing package with its documentation unchanged. Had the developer known, they could have changed the import or installed the named package instead to bug-solve, but developers often consider package documentation and examples important in evaluating package quality}. The developer uninstalls the package using the command \texttt{npm uninstall node-create-csv} and decides to look \hl{at the other packages NPM returned, which may have working examples.}

\begin{figure}[h]
    \begin{lstlisting}[language=diff]
let converter = require('json-2-csv');
+ const fs = require('fs');
- let documents = [{Make: 'Nissan', ... },
+ let documents = [{Name: 'Alice', 
+       Institution: 'Foo', Job: 'IT Manager' },
-   {Make: 'BMW', ...}];
+   {Name: 'Bob', Institution: 'Bar',
+       Job: 'Developer'}];
let json2csvCallback = function (err, csv) {
    if (err) throw err;
    console.log(csv);
+   fs.writeFile('./people.csv',csv,function(err) {
+       if(err) return console.log(err) })
};
converter.json2csv(documents, json2csvCallback);
    \end{lstlisting}
    \caption{The modifications the developer made to complete the task, with new lines marked in green with ``+'' and removed lines in red with ``-''.}
    \label{fig:manualexample}
\end{figure}

\hl{The developer evaluates another three packages from their search results, reading their documentation, however none have any examples in their README files. Finally, the developer checks the} \texttt{json-2-csv} package. In the README for this package, there are no usage examples, but a link to some examples on GitHub, where they find one example for writing a CSV file. They install the package and paste the example in their editor. 

Figure~\ref{fig:manualexample} shows how the example code for this package was adapted to solve the assigned task. The original code snippet takes an array of data, \CodeIn{documents}, and defines a callback function where the argument \CodeIn{csv} is printed to the console. The developer modifies it so the array contains their data, and in the callback, the string is written to a file called ``people.csv''. They run this code and the file is created successfully. In total, this participant took 18 minutes to complete this task.

To sum up, this approach can be time-consuming for the following reasons:
\begin{itemize}[leftmargin=*]
    \item Developers need to search different sources to locate potentially useful packages. When finding those package, developers need to read documentation and check examples, when available;
    \item \hl{Not all packages have examples, making evaluating package quality more difficult.}
    \item The package of interest may \emph{not} work as expected. Running code snippets is important to make a decision on the package to use. For that, the user needs to create temporary environments and (un)install packages on top of those environments to test different packages.
\end{itemize}

\subsection{Problem solving in \tname}
\label{sec:automatedexample}

This section illustrates how developers solve tasks using \tname{}, a tool to help developers search for Node.js packages and experiment with related examples. We describe the workflow of another developer in our user study. The developer was asked to solve the same task from Section~\ref{sec:manualexample}.

\begin{figure}[h]
    \centering
    \includegraphics[width=0.9\linewidth]{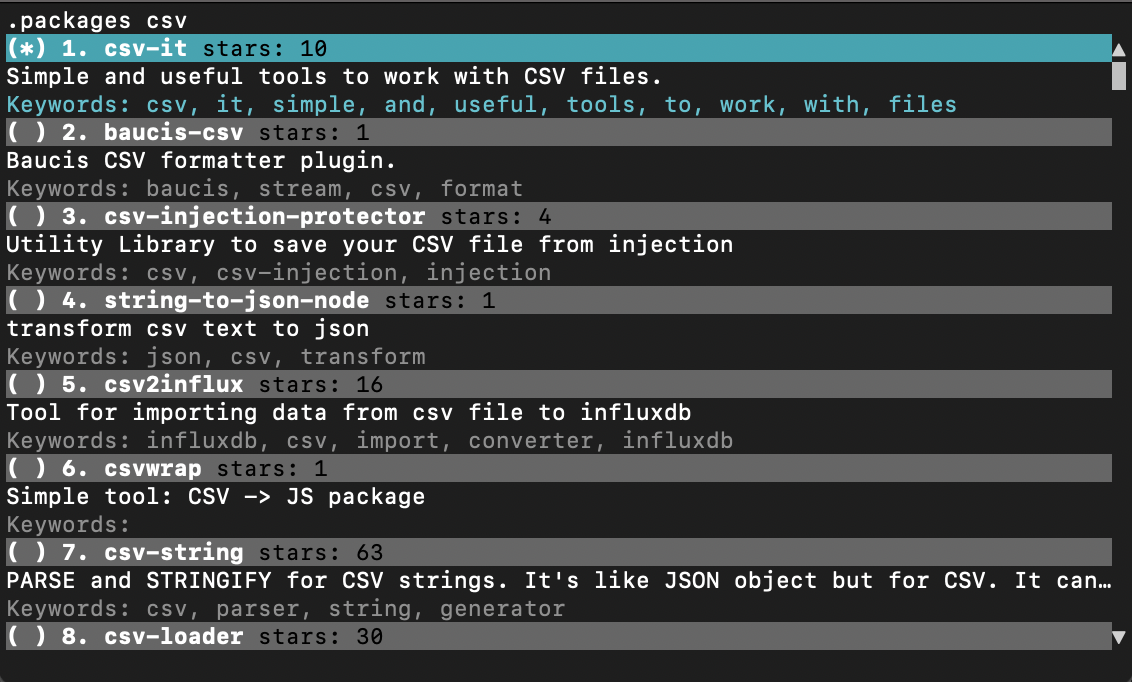}
    \caption{Results for the query `.packages csv'.}
    \label{fig:packages}
\end{figure}

First, the developer starts the tool and uses the \texttt{repl} command to create a new Read-Eval-Print-Loop (REPL) instance, where they can install packages and execute code. The developer then uses the \texttt{.packages <search string>} command from within the REPL to search for packages using the search string. This command takes the developer to a list of packages and their descriptions, which they can scroll through. Figure~\ref{fig:packages} shows the output the REPL produces for the command \texttt{.packages csv}.

The list of packages allows for skimming many descriptions to quickly disregard irrelevant packages, for example, \texttt{csv2influx}, which appears to import CSV files into a database. \tname{} sorts packages based on their runnability, and excludes packages with no stars or no example code in their READMEs. The first entry, \texttt{csv-it}, looks promising so they select it with the enter key. Upon selection, the tool asks if they would like to install the package, to which they respond ``yes''.

\begin{figure}[h]
    \centering
    \includegraphics[width=0.9\linewidth]{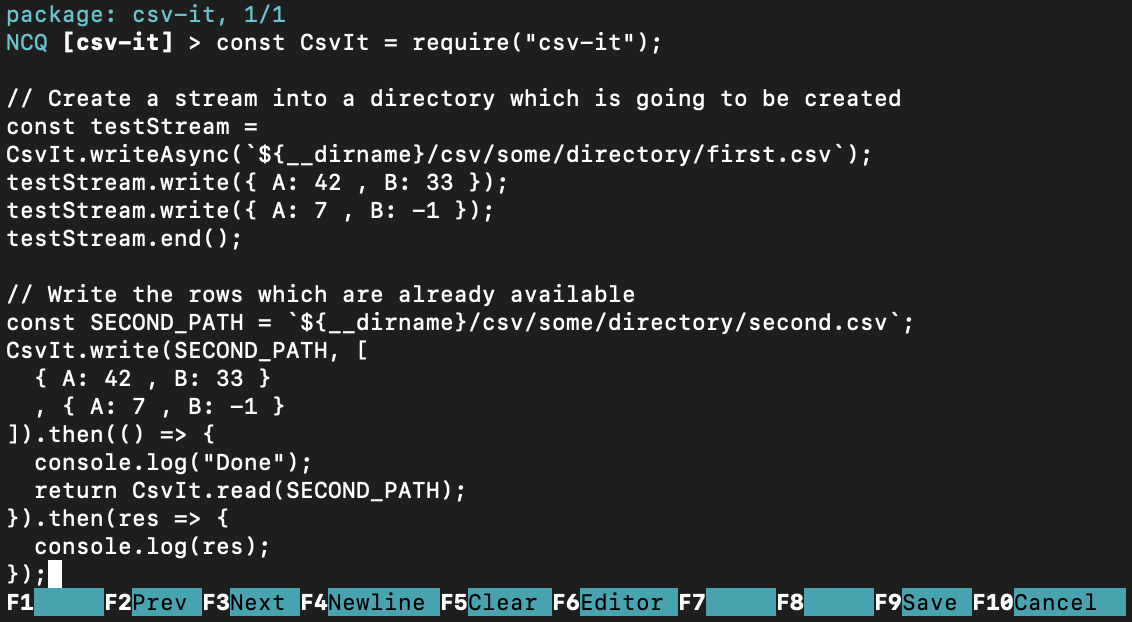}
    \caption{Cycling through code snippets for the package \texttt{csv-it} using the command \CodeIn{.samples}.}
    \label{fig:samples}
\end{figure}

To evaluate a package, developers can use the \texttt{.samples <package name>} command to retrieve code snippets for a given package, extracted from NPM by scrapping the README file associated with a package. The developer uses the command \texttt{.samples} to see examples for the installed packages. Figure~\ref{fig:samples} shows the code snippet for \texttt{csv-it}. The code snippet autofills the command prompt; from here the developer can execute the code snippet like any other code in the REPL by pressing enter, or cycle through other code snippets by using the function keys F2 and F3. Each code snippet is evaluated using ESLint to correct errors and run through \tname{}'s error correction process. This process fixes errors, enforces consistent style rules on snippets and comments out error-causing lines, before being sorted by final error count to show error-free snippets first.

\begin{figure}[h]
    \centering
    \begin{lstlisting}[language=diff]
const CsvIt = require("csv-it");
- const testStream = CsvIt.writeAsync(`first.csv`);
+ const testStream = CsvIt.writeAsync(`people.csv`);
- testStream.write({ A: 42 , B: 33 });
+ testStream.write({ Name: "Alice",
+   Institution: "Foo",  Job: "IT Manager"});
- testStream.write({ A: 7 , B: -1 });
+ testStream.write({ Name: "Bob", 
+   Institution: "Bar", Job: "Developer"});
testStream.end();
- const SECOND_PATH = `second.csv`;
- CsvIt.write(SECOND_PATH, [{ A: 42 , B: 33 } , 
-   { A: 7 , B: -1 }]).then(() => {
-  console.log("Done");
-  return CsvIt.read(SECOND_PATH);
- }).then(res => {console.log(res);}); 
    \end{lstlisting}
    \caption{How the code snippet in Figure~\ref{fig:samples} was adapted.}
    \label{fig:autoexample}
\end{figure}

In this case, there is only one code snippet, which demonstrates how to write some data to a CSV file in two different ways. \tname{}'s automated fixes have added semi-colons where they are missing; Figure~\ref{fig:autoexample} shows the changes the developer makes to the code snippet to complete their task. First, the developer removes the second example. Next, they decide to run the code snippet by pressing enter and see that it works as expected; a file named ``first.csv'' is created with the columns ``A'' and ``B'' and two rows of values. After executing, the developer uses the \texttt{.editor} command, which allows them to edit their previously run code in a traditional editor, where they change the file name and data before saving. On save, the REPL re-runs the code successfully. In total, this participant took 15 minutes to complete the task.

In summary, \tname{} \hl{combines ideas from existing work in code and package search to} aid code reuse by:
\begin{itemize}[leftmargin=*]
\item Automatically setting up temporary environments that developers can use to experiment with packages;
\item Allowing developers to search for packages without leaving their programming environment~\cite{10.1145/3368089.3417922}; \hl{developers only see packages with example code, making the package to code process faster};
\item \hl{Sorting packages by runnability to prioritise functioning packages; the majority of developers in our user study were able to solve their task with the first package they found;}
\item \hl{Embedding code snippets within the editor}~\cite{Campbell2017NLP2CodeCS}\hl{ and enabling developers to quickly try out different packages without leaving the tool, while providing immediate error feedback when code is submitted, via the use of the REPL design. Developers are easily able to check if code snippets work before making changes;}
\item The addition of a traditional editor mode so changes can be made to the REPL's state, which makes the REPL more intuitive for users;
\item Automatically fixing errors and sorting snippets by number of errors to reduce the amount of work developers do to reuse code~\cite{NLP2TestableCode}.
\end{itemize}


\section{Dataset}
\label{sec:datacollection}

\tname{} uses a database of packages, package data and code snippets to enable its search. Like previous work~\cite{NLP2TestableCode}, we implement this database offline to avoid internet bandwidth issues. The following sections describe how package data and code snippets were mined. Table~\ref{tab:dataset} shows the package data breakdown; our final dataset consists of 620,221 packages with 2,161,911 code snippets.

\begin{table}[h]
    \centering
    \begin{tabular}{lr}
         Packages as of May 2021 &  1,607,057 \\
         Packages with READMEs & 1,298,170 \\
         Packages with non-empty READMEs & 1,297,678 \\
         \midrule
         Packages with Node.js code snippets & 620,221 \\
         \multicolumn{2}{c}{}\\
         Node.js code snippets & 2,161,911
    \end{tabular}
    \caption{Summary of dataset statistics.}
    \label{tab:dataset}
\end{table}
\vspace{-0.4cm}

\subsection{Package Data} 

To enable the package and code search, we mined the NPM registry for NPM package data. 
\hl{We chose to restrict our focus to a single package manager to simplify automation. Many packages are also hosted across multiple package managers, so focusing on a single one limits data duplication and the size of our database. We selected NPM due to it being one of the most popular package managers for Node.js}.
The registry maintains a list of all package names at the URL \url{https://replicate.npmjs.com/_all_docs}.  For each package, a JSON file containing package data (\eg{}, README, description, author and link to the repository) can be found at the URL \url{https://registry.npmjs.org/<package>}.

There were 1,607,057 packages on the NPM registry as of May 2021. We were able to mine the name, descriptions, keywords, README, last modification date and repository URL for 1,298,170 packages. As NPM has a 64KB limit on README length on the registry, we also downloaded the full README from corresponding GitHub project, when linked. For 308,887 packages (19.2\% of all packages), there was no README available on either site. Next, we excluded empty README files, leaving us with 1,297,678 non-empty READMEs. Finally, we limited our dataset to the 620,221 packages with at least one code snippet, following the process in Section~\ref{sec:cs_extraction}.

As the NPM registry does not contain repository data relating to package popularity, we mined GitHub for additional package data, such as the number of stars and if there was a license. We then processed the README to count number of lines, markdown code blocks of any language and if there were install or run examples. We determined if there was a run or install example by looking for the headings ``install'' or ``usage'', or the use of the commands \CodeIn{npm run} or \CodeIn{npm install}.

\subsection{Code Snippet Extraction}
\label{sec:cs_extraction}

Code snippets for the code search were extracted from README files. We limited our focus to README files over other documentation as this was the easiest to systematically mine. It is worth mentioning that NPM recommends all packages include a README (and consequently 80.8\% of packages on the registry have a README). Other forms of software documentation do exist, but differ across projects, making it harder to mine. We also chose not to mine repository code for snippets as downloading the source of every NPM package would be very time intensive.

To extract code snippets, we looked for code blocks in the README markdown. These code blocks often include language information for syntax highlighting, which we used to discard code snippets marked as a non-\javascript{} language (e.g., bash commands used to demonstrate package installation or JSON for results of execution). However, as not all code snippets use syntax highlighting, we also analysed the remaining dataset to further filter non-\javascript{} snippets. We excluded snippets starting with common commands (\CodeIn{npm install} for example), and used a regular expression to look for JSON objects.

As code snippet extraction is an automated process based on heuristics, we manually verified the results using a statistically representative random sample of 384 READMEs (confidence level of 95\% and a confidence interval of 5). Two authors manually compared each README and the resulting code snippets from our extraction process, checking for any missing or non-\javascript{} snippets. \hl{The authors had a Cohen's Kappa agreement of 0.72. Then, the authors met to discuss results and determine a final categorisations. Using this data, we refined the extraction process to handle any erroneous cases and reviewed again to verify. Ultimately, we were able to extract 2,161,911 code snippets from the 620,221 READMEs.}

\section{Survey}
\label{sec:survey}

As a preliminary step in the design \tname, we conducted a survey to understand how developers find NPM packages and what challenges they experience. The following sections describe our survey design and the results.

\subsection{Survey Design}

To find Node.js programmers, we recruited participants from Prolific\footnote{\url{https://www.prolific.co/}} --a platform to recruit research participants-- and employed a screening survey using a series of programming knowledge questions adapted from Danilova \etal{}~\cite{doyoucode}. \hl{Questions were designed to have a single `correct' answer, so that they could easily be used to verify if the participant had basic programming knowledge.} The full details of our screening, \hl{including questions and answers}, are available in a workshop paper~\cite{prolificUserStudy}. 

\begin{table}[h]
    \centering
    \small
    \caption{Survey Questions (Q)}
  \rowcolors{3}{white}{LightGray}
    \begin{tabularx}{\linewidth}{lX}
    \toprule
    \hiderowcolors 
    \showrowcolors
    Q1 & How often do you Search for NPM packages? \\
    Q2 & What site do you use the most when searching for NPM packages? \\
    Q3 & What kind of challenges do you have searching for NPM packages? \\
    Q4 & What code editor do you use the most to program in Node.js? \\
    \bottomrule
    \end{tabularx}
    \label{table:surveyq}
\end{table}

Table \ref{table:surveyq} shows the set of questions we asked participants. Questions were selected to give insights into the frequency of NPM package use, websites used to find packages and the challenges developers experience. In addition, to understand the way developers program in Node.js in general, we asked about code editor usage (Q4). 

We evaluated the \textit{credibility} of participants using the framework proposed by Rainer and Wohlin \cite{framework}. We aimed to recruit `performers', in this case, software developers with real, relevant and recent experience according to the R\textsuperscript{3} model. We used the programming skill questions to verify \textit{real} and \textit{relevant} Node.js programming skill, and participants were asked how often they programmed in Node.js to assess that the experience was \textit{recent}. Furthermore, we used Q1 to again assess relevancy of experience in regards to package search; participants that answered `Never' were not shown Q2 and Q3.

We received 680 responses, out of which 55 answered all questions correctly and were determined to be Node.js programmers.

\subsection{Survey Results}

\begin{figure}[h]
    \centering
    \includegraphics[width=0.95\linewidth]{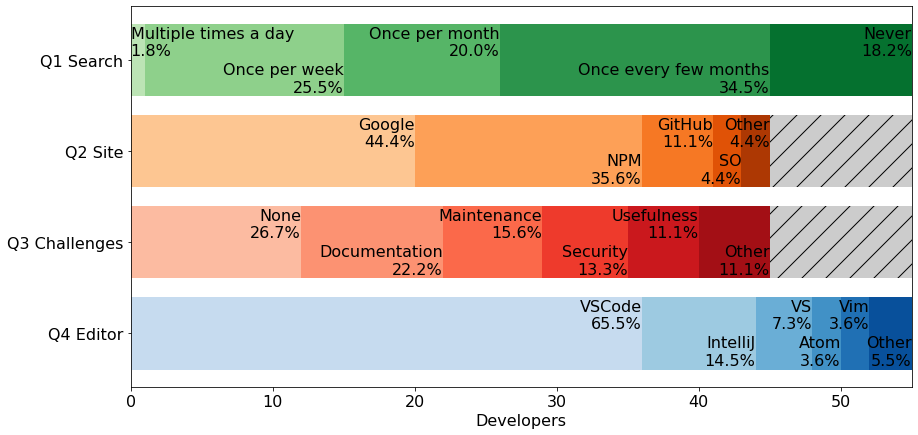}
    \caption{Responses to survey questions for 55 developers. Q2 and Q3 record no response for developers who do not search for NPM packages.}
    \label{fig:surveyResponses}
\end{figure} 

Figure~\ref{fig:surveyResponses} shows the results of our survey. We found that out of 55 developers, 45 searched for NPM packages~(Q1) once every few months or more often (81.8\%). Considering these developers, the most popular websites to search for packages were Google (44.4\%) and the NPM website (35.6\%), with most developers (73.3\%) reporting that they experienced challenges. The most common challenges mentioned were insufficient documentation and example code (22.2\%), followed by maintenance (15.6\%) and security issues (13.3\%). Developers highlighted issues finding packages outside the most popular, with one developer saying ``Maybe the biggest grip[e] is trying to find packages outside of what's popular that are well documented''. In general, most developers had issues finding high quality packages with so many options on the NPM registry; as stated by one developer ``I also have a hard time finding out what all the alternatives are to see what my options are''. The results for Q4 show that VSCode was the most popular editor for Node.js within our sample. 65.5\% of developers answered VSCode, followed by IntelliJ at 14.5\%. Therefore, we use it as a baseline for the study described in Section~\ref{sec:eval2}.

In summary, package use was a common part of programming in Node.js, but most developers had challenges finding high quality NPM packages.

\section{\hl{Overview}}
\label{sec:overview}

\begin{table}[h!]
    \centering
    \begin{tabular}{ll}
        \toprule
         \multicolumn{1}{c}{Tool} & \multicolumn{1}{c}{Baseline} \\
        \midrule
        \tnamei{} & \baseline{} + Web \\
        \tnameii{} & VSCode + Web \\
        \bottomrule
    \end{tabular}
    \caption{Comparison of baselines used in each evaluation.}
    \label{table:ncqversions}
\end{table}

\hl{To address the problem described in Section}~\ref{sec:manualexample}\hl{, we developed Node Code Query (NCQ). \tname{} is a command line tool for Node.js that integrates code and package search into a REPL development environment. The tool extends the existing Node.js REPL and, along with the ability to execute code, adds the ability to search for NPM packages using natural language queries, find and run example code snippets and install and uninstall packages, all without leaving the environment. We first developed an initial version of \tname{}, \tnamei{}, which focuses on retrieving packages and code snippets, then, using feedback from a user study of 10 developers, we refined the tool further, adding additional features such as better package search and code correction.}

\hl{Table}~\ref{table:ncqversions} \hl{shows each version of the tool and the baseline used to evaluate against. For the first user study, we evaluate \tnamei{} against a baseline version of the tool with no search features and web access; we justify this decision as evaluating the impact of \tnamei{}'s search features. For the second evaluation, we then compare \tnameii{} against a more realistic baseline of VSCode and web access, to see how the tool performs against `typical' processes. We do not evaluate the two variants of \tname{} against each other as this research focuses on the comparison between automated and manual code reuse techniques. The variants are not fundamentally different besides the additional features, described in Section}~\ref{sec:ncq2}.

\section{\tnamei{}}
\label{sec:ncq1}


\begin{figure}[h]
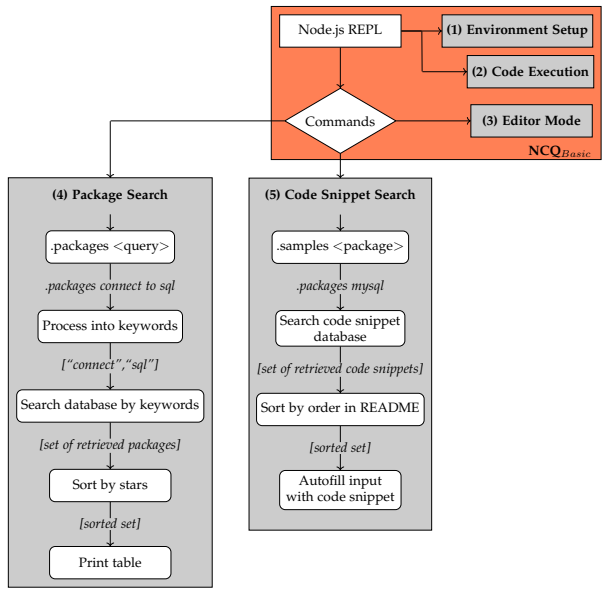

\centering
\includestandalone[width=0.9\linewidth]{figures/ncqrfeatures}
\caption{Overview of \tnamei{}'s features. Features shared with \baseline{} are highlighted in orange.}
\label{fig:pipeline}
\end{figure}

The initial version of \tname{}, \tnamei{}, focuses on the problem of finding packages and code snippets. As finding and testing packages is often a process with many small tasks where developers may want to quickly try out different packages, we implement \tnamei{} as a \hl{Read-Eval-Print-Loop (REPL)}. The REPL model is highly suited to this type of programming; developers receive immediate feedback on execution of each line, and writing and executing code are tied together. Figure~\ref{fig:pipeline} gives an overview of the features in \tnamei{}; \hl{1) the ability to setup environments, 2) code execution, 3) the editor mode, 4) package search and 5) code snippet search.}

\subsection{REPL}
Starting the REPL creates a new environment to execute Node.js code and install packages. The tool creates a new directory and the necessary project files that allow a user to install packages: (1)~the \texttt{package.json} file, which stores project details and directly installed packages, and (2)~the \texttt{package-lock.json}, which lists the entire dependency tree. These files are automatically updated by the package manager when installing/uninstalling packages. 

\subsection{Editor Mode}
To make programming in a REPL easier, \tnamei{} includes an additional editor mode that allows users to edit the state of the REPL in a traditional editor. The default Node.js REPL maintains a context of executed code as users submit code, but this cannot be modified besides overwriting previously defined functions and variables. This often means that users need to rewrite entire functions to fix errors. \tnamei{}'s editor mode provides a simple text editor that presents a user with their previously run code, where they can edit or remove lines. On saving the file, the REPL state will be reset and the new code will be run.

\subsection{Package Search}
\tname{}'s package search replies on a database of NPM package information, as described in Section~\ref{sec:datacollection}. Packages are pre-processed to create a set of keywords they are indexed by. First, package descriptions are separated into unique words, then combined with the package's keyword field, processed to remove stop words (common, individually meaningless words like ``the'' and ``to''). Then, the remaining set of words are stemmed using the Porter Stemmer algorithm.\footnote{\url{https://tartarus.org/martin/PorterStemmer/}}

\begin{figure}[h]
    \centering
    \includegraphics[width=0.9\linewidth]{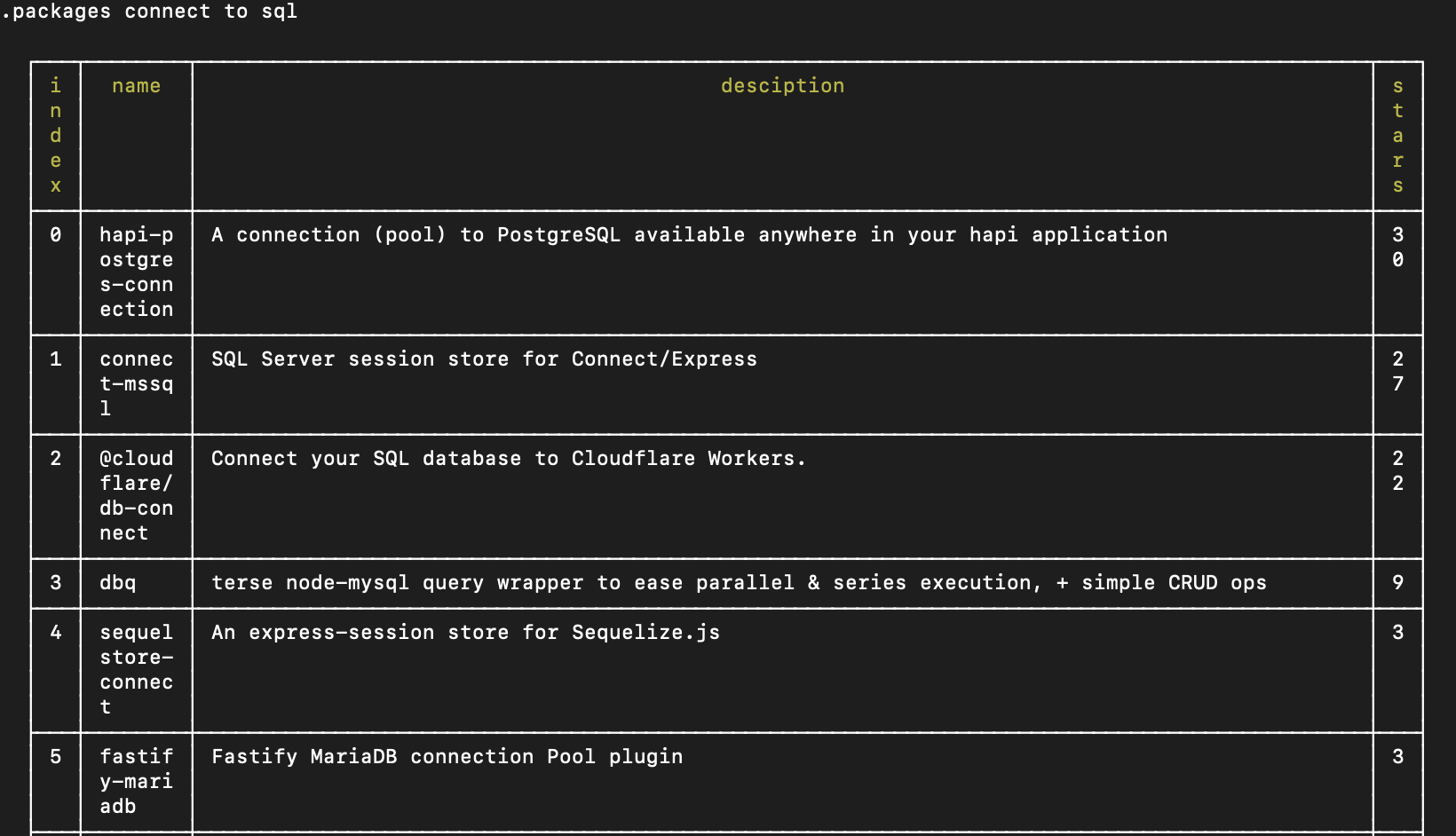}
    \caption{Output of the \texttt{.packages} command.}
    \label{fig:package_table}
\end{figure}

A similar process is taken for queries using the \texttt{.packages} command, which Figure~\ref{fig:pipeline} illustrates. For example, the query ``connecting to sql'' is first separated into the words ``connecting'', ``to'', and ``sql'', then stop words are removed, leaving only ``connecting'' and ``sql''. After stemming, the set of words is ``connect'' and ``sql''. For each of these words, a set of associated packages can be retrieved; the final result is the subset of packages present in all sets, which is sorted by stars then tabulated with information such as package stars and descriptions to help users make a choice (Figure~\ref{fig:package_table}).

\subsection{Code Snippet Search}
\tnamei{} includes a code snippet search that allows developers to look for example code snippets for a package. Each snippet in our dataset has an associated package it originates from and an index based on the order of appearance in the original README.

As shown in Figure~\ref{fig:pipeline}, using the \texttt{.samples} command, code snippets belonging to the given package are returned in order of their occurrence in the original README. Snippets autofill the command line and can be modified or run without needing to copy and paste. We preserve the original order of code snippets to enable developers to more easily understand the logical order some snippets are intended to be executed in; not all code snippets run on their own and they often leave out steps shown in previous snippets, for example, importing the package. Related code snippets also remain grouped together due to this order.


\section{\tnamei{} Evaluation}
\label{sec:eval1}


To evaluate \tnamei{} we conducted a user study with 10 Node.js developers, where participants completed programming tasks using the tool. We pose the following related research questions:

\newcommand{\rqone}{What is the influence of \tnamei{}'s search features on the performance of participants?}
\newcommand{\rqtwo}{What are the perceptions of participants about \tnamei{}?}
\newcommand{\rqthree}{What is the influence of \tnameii's search features on the performance of participants?}
\newcommand{\rqfour}{What are the participants' perceptions about \tnameii{}?}
\newcommand{\rqfive}{How effective is \tnameii{}'s code correction?}

\begin{adjustwidth}{0em}{0pt}
\vspace{1ex}
\noindent
\textbf{RQ1.} \rqone{}

\noindent
\textbf{RQ2.} \rqtwo{}
\vspace{1ex}
\end{adjustwidth}

\noindent
The purpose of \textbf{RQ1} is to understand if the tool's search features are helpful, using the comparison baseline of a version of \tname{} that lacks search features. Instead, participants used the internet to find packages and examples. We used the following metrics to measure the impact of \tnamei{}'s search:~1)~time to completion, 2)~how quickly developers found and installed packages and 3)~ the number of packages developers tried. The purpose of \textbf{RQ2} is to understand developer perceptions of the tool. We analysed participant responses to a set of questions we asked about their experiences after each task.

\subsection{Experimental Design}
\label{sec:experimental-design}

This section describes the design of the user study, including how we selected the tasks, what baseline we used, how we recruited participants, how we assigned tasks to participants, how the sessions were run and what questions were asked to participants.

\subsubsection{Programming Tasks}
\label{sec:tasks}

As the user study focuses on solving tasks, task selection was important. To represent real programming tasks, we looked at a list\footnote{\label{fn:tasks}\url{https://bit.ly/2HK5tLq}} of mined Stack Overflow questions from previous work on the NLP2TestableCode project~\cite{NLP2TestableCode}. We looked for examples that could be solved in 20 minutes or less by a developer with basic experience in Node.js. It is important to note that, to facilitate the checking of the solutions provided by participants, we modified the original tasks to specify example inputs. For example, to complete task~4, we provide as input to the participant the SQL database to be used so that we could validate the output.

We used the following five tasks:

\begin{enumerate}[leftmargin=*]
    \item Read a text file (any) from the file system to memory, using a package other than ‘fs’. (Sources: Line 9893 of the task list and \stackoverflow{} question \href{https://stackoverflow.com/questions/20186458}{20186458}.)
    \item Create an array with ten randomly-chosen numbers and sort the array with ``merge sort''. (Sources: Line 585 and 99292 and SO questions \href{https://stackoverflow.com/questions/42033229}{42033229} and \href{https://stackoverflow.com/questions/2571049}{2571049}.)
    \item Use a package to create the CSV file \CodeIn{people.csv} with the contents: 
\end{enumerate}
\begin{Verbatim}[fontsize=\small,xleftmargin=.2in]
  Name, Institution, Job
  Alice, Foo, IT Manager
  Bob, Bar, Developer
\end{Verbatim}
\begin{adjustwidth}{2em}{0pt}
(Sources: Line 3073 and SO question \href{https://stackoverflow.com/questions/59600762}{59600762}.)
\end{adjustwidth}

\renewenvironment{quote}{%
  \list{}{%
     \leftmargin0.5cm   
     \rightmargin\leftmargin
  }
  \item\relax
}
{\endlist}

\begin{enumerate}[leftmargin=*]
  \setcounter{enumi}{3}
    \item Connect to a MySQL database that we provide and retrieve customers using the SQL command \CodeIn{"SELECT customerName FROM customers WHERE country = `Australia';"}.
    (Sources: Line 221 and SO question \href{https://stackoverflow.com/questions/6597493}{6597493}.)
    
    \item Create a directed graph in memory with the following edges \{(a,b), (a,c), (b,c), (c,d)\}. The vertex ``a'' is the root of the graph. Print the vertices on the console as they are visited in depth-first order. Print each vertex only once. (Sources: Line 13232 and SO question \href{https://stackoverflow.com/questions/14483473}{14483473}.)
\end{enumerate}

\hl{
Participants were said to have successfully completed a task if they produced a \javascript{} file that, when run, satisfied the task description. Tasks were designed so that it would be straightforward to determine completion.
}

\subsubsection{Baseline}
\label{sec:baseline1}

We designed the baseline of this evaluation as \baseline{}, a version of \tnamei{} with package and code search suppressed. \baseline{}'s features are highlighted in orange in Figure~\ref{fig:pipeline}. With \baseline{}, participants must search for packages online, using the browser. \hl{We put no limits on what sites could be used for search, instead of restricting them to just the NPM website, as the results of our survey indicate that developers prefer to use generic search engines like Google to find packages.} Participants first solve a task using the baseline, then using \tnamei{}, in order to compare the effect of search on results while reducing the impact of tool familiarity. We expect that developers would be more familiar programming in a traditional editor and that this could have an impact on results.

\subsubsection{Participants}
\label{sec:participants}

We hired \numparticipants{} developers from online, for example, using social media sites such as Twitter and Telegram. We offered participants \$40USD and required fluency in English and basic knowledge in Node.js. We estimated the total time of a user session as follows: 10 minutes for presentation/training, 20 minutes to run a task with \baseline, 20 minutes to run a task with \tnamei{} and 10 minutes to answer questions after tasks (1hr total). Participants had an average of 2 years experience with Node.js.

\subsubsection{Assignment}
\label{sec:task-assignment}
\begin{table}[h!]
    \centering
    \caption{\label{table:taskassign}Fragment of task assignments.}
    \begin{tabular}{cccccc}
        \toprule
        \multirow{2}{*}{participant} & \multicolumn{5}{c}{task}\\
        & a & b & c & d & e \\
        \midrule
        1 & N & Y & - & - & - \\
        2 & - & N & Y & - & - \\
        3 & - & - & N & Y & - \\
        4 & - & - & - & N & Y  \\
        5 & Y & - & - & - & N  \\
        \bottomrule
    \end{tabular}
\end{table}

We used a typical combinatorial design to uniformly assign tasks to participants. For space reasons, Table~\ref{table:taskassign} shows a fragment of the design including only 5 participants (rows) and 5 tasks (columns). A cell with the label ``Y'' (respectively, ``N'') indicates that a participant will be assigned to solve a given task with \tnamei~(respectively, \baseline{}). Note that every participant executes one task with \tnamei\ and one task with the baseline and, in this design with 5 participants, every task is executed by two participants. The design of this table is a Latin Square as the number of participants and tasks are the same, but the actual design we used included \numparticipants\ participants to obtain a larger sample. Intuitively, that design can be obtained by repeating the rows associated to participants 1-5 two times. With \numparticipants\ participants, each task is executed by 4 different participants: 2 participants using \tnamei\ and 2 participants using \baseline{}. 

\subsubsection{User Session}
\label{sec:user-session}

The user study was undertaken remotely using AnyDesk\footnote{\url{https://anydesk.com/en}}, with participants connecting at an agreed upon time with the instructor (first author of this paper). Before the session, participants were asked to watch a video demonstration of the tool\footnote{\url{https://youtu.be/C1PZ2g96eVo}}. At the beginning of the session, the instructor demonstrated the features of each tool and then the participant attempted each task. After the participant turned in each task, the instructor asked the participant to complete a section of the questionnaire (see Section~\ref{sec:questions}).

Tool documentation (\eg{}, keys and commands) was made available to participants on the right side of the screen, with the tool itself on the left. Experiments were run within a virtual machine with the necessary infrastructure to run each task provided; for example, for task~4, the database was already set up and a file was provided with the connection details. Snapshots were used to restore the environment to the same initial state for each participant.

Audio and video (screen capture) were recorded for each session. The tool was configured to record timestamped input (use of commands and submitted code) into log files. At the end of each task the user's REPL code was saved into a file. An additional log of visited websites was transcribed from the session recordings.

\subsubsection{Questions}
\label{sec:questions}

\begin{table}[h]
    \centering
    \caption{
    Questions part A, B and C.
    }
  \rowcolors{3}{white}{LightGray}
    \begin{tabularx}{\linewidth}{lX}
    \toprule
    \hiderowcolors 
    \showrowcolors
    A1 & How many years of experience do you have with Node.js? \\
    B1/C1 & Overall, did you consider the tool helpful to accomplish the assigned task? \\
    B2/C2 &  Grade your confidence, in a 1-7 scale, that your solution is correct according to the problem specification? \\
    C3 & Did you consider the command ``packages'' useful? \\
    C4 & Did you consider the command ``samples'' useful? \\
    C5 & What features did you like about the tool? \\
    C6 & What you did not like about the tool? \\
    \bottomrule
    \end{tabularx}
    \label{table:questionsABC}
\end{table}

Participants were asked to complete sections of a questionnaire via a Google Form at the start of their session (Questions A), after the first task (Questions B) and after the second task (Questions C). Table \ref{table:questionsABC} shows the set of questions labeled by stage; for questions B1-2 and C1-4 participants were asked to rank agreement on a seven-level Likert scale. 

\begin{table}[h]
    \centering
    \caption{Features provided to users as answers (F) for C5.}
  \rowcolors{3}{white}{LightGray}
    \begin{tabularx}{\linewidth}{lX}
    \toprule
    \hiderowcolors 
    \showrowcolors
    F1 & Ability to execute code through REPL\\
    F2 & Ability to search for packages\\
    F3 & Ability to search for code snippets\\
    F4 & Ability of using an editor to modify snippets (without leaving the REPL) \\
    F5 & The combination of code search and execution (virtualization) in a single tool \\
    F6 & Ability to install and uninstall packages in an environment \\
    F7 & Ability to save work and resume from outside the tool (.save) \\
    F8 & Ability to access main functionalities through function keys \\
    \bottomrule
    \end{tabularx}
    \label{table:features}
\end{table}

For each task we asked participants about their confidence in their solutions (B1 \& C1) and the helpfulness of the tool they used (B2 \& C2). Additionally, in Section C participants were asked about \tnamei{}'s features, for example, if they found certain commands useful (C3 \& C4), what features they liked (C5) and what they didn't like (C6). For Question C5, participants were given the list of features seen in Table~\ref{table:features} and allowed to select multiple options. The questionnaire concluded with an option to provide suggestions.

\begin{figure*}[t]
\begin{center}
\includegraphics[width=0.95\linewidth]{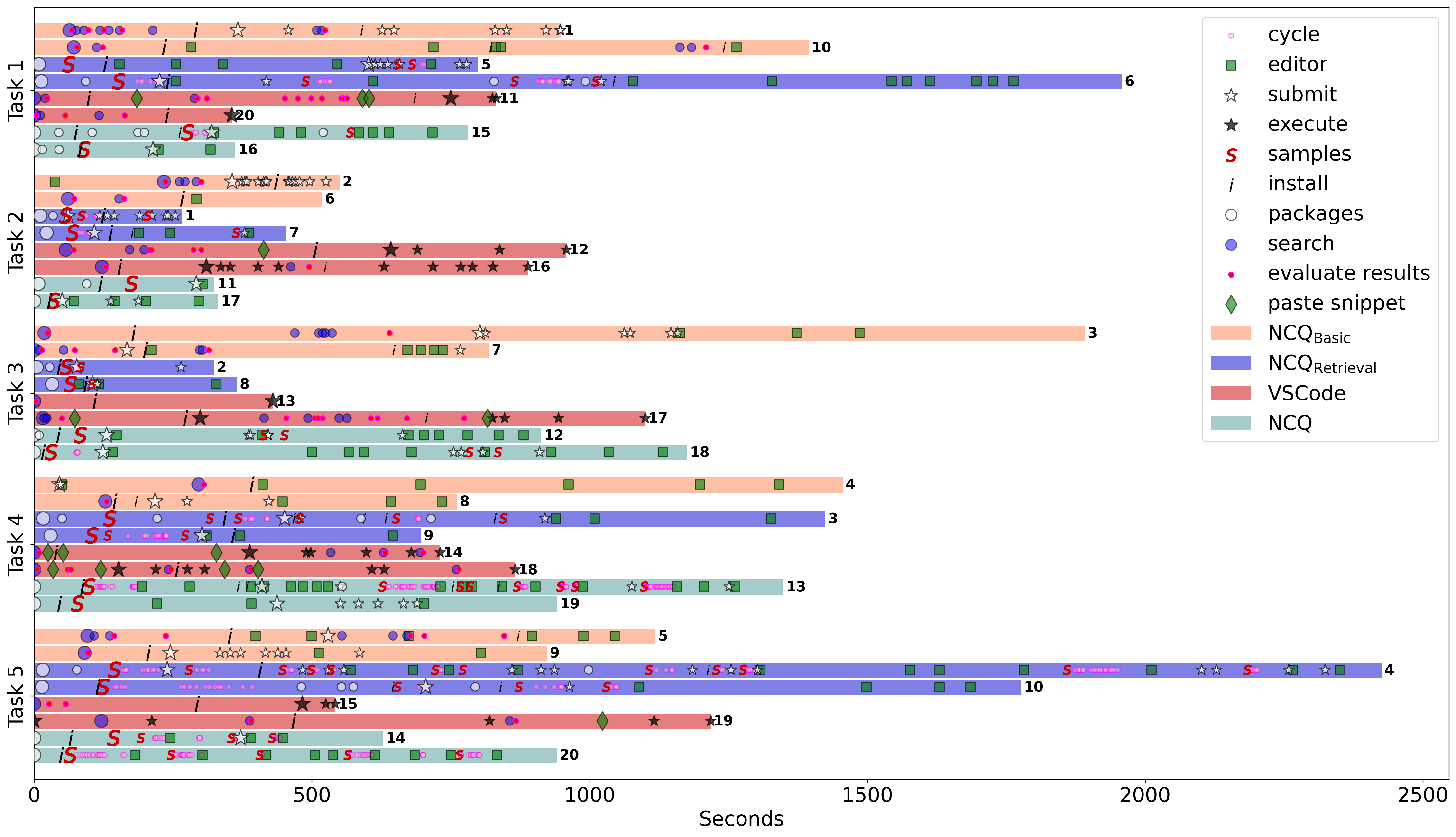}
\caption{Timeline of all participants across both user studies, grouped by task and treatment. The $x$-axis shows time in seconds, markers show features used over time, with the first usage enlarged. Each bar is labeled by participant ID.}
\label{fig:timeline}
\end{center}
\vspace{-1.5em}
\end{figure*}

\begin{table}[h!]
    \centering
    \caption{TAM (Technology Acceptance Model) questions (D).}
    \rowcolors{3}{white}{LightGray}
    \begin{tabularx}{\linewidth}{lX}
    \toprule
    \hiderowcolors 
    \showrowcolors
    D1 & Using the tool would improve my performance in my job. \\
    D2 & Using the tool in my job would increase my productivity. \\
    D3 & Using the tool would enhance my effectiveness in my job. \\
    D4 & I would find the tool to be useful in my job. \\
    D5 & My interaction with the tool is clear and understandable. \\
    D6 & Interacting with the tool does not require a lot of my mental effort. \\
    D7 & I find the tool to be easy to use. \\
    D8 & I find it easy to get the tool to do what I want it to do. \\
    D9 & The quality of the output I get from the tool is high. \\
    D10 & I have no problem with the quality of the tool's output. \\
    D11 & I would have no difficulty telling others about the results of using the tool. \\
    D12 & I believe I could communicate to others the consequences of using the tool. \\
    D13 & The results of using the system are apparent to me. \\
    D14 & I would have difficulty explaining why using the system may or may not be beneficial. \\
    \bottomrule
    \end{tabularx}
    \label{table:questionsD}
\end{table}

At the end of the session participants answered questions based on the Technology Acceptance Model (TAM)~\cite{TAMdavis1989perceived}, a model for measuring, predicting, and explaining use of technology, including software. Our questions were adapted from TAM2~\cite{TAMvenkatesh2000theoretical}, which also incorporates job relevance and output quality. These questions can be seen in Table \ref{table:questionsD}.

\subsection{Answering RQ1:~\rqone{}}

Prior research has found that context switching (e.g., switching from a programming environment to a web browser) can disrupt productivity~\cite{Proksch2015}. In Section~\ref{sec:baseline1} we describe a version of \tnamei{} with no search features; instead, participants search resources online. We compare this to \tnamei{}, with the hypothesis that the integrated search features will impact participant performance. The following sections elaborate on the research questions (associated with RQ1) that we posed.


\begin{figure}[h!]
\begin{center}
\includegraphics[width=0.96\linewidth]{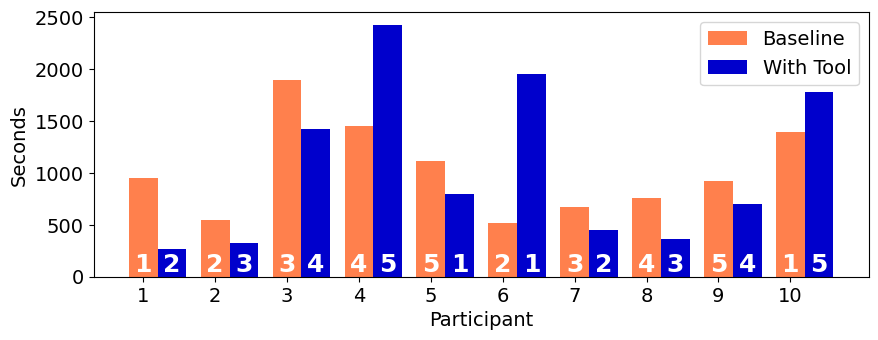}
\includegraphics[width=0.96\linewidth]{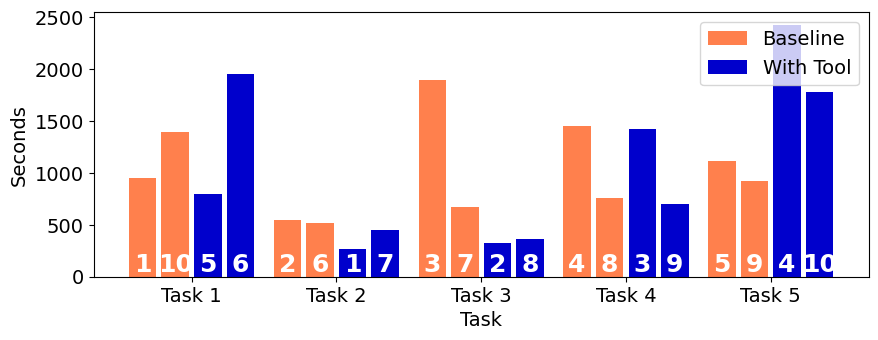}
\caption{\hl{Duration of tasks by participant and by task. Bars are labelled by task ID in the first figure, and by participant ID in the second figure.}
}
\label{fig:taskdurations}
\end{center}
\end{figure}

\subsubsection{Was there impact on time to complete each task?}
\label{sub:impact-on-time}


Figure~\ref{fig:taskdurations} shows duration to solve tasks (in seconds), by participant and by task, using \baseline\ (left) and \tnamei\ (right). \hl{All participants solved their tasks.} We compare the difference in durations for each participant, and find that 7 out of 10 participants completed tasks faster when using \tnamei\ instead of \baseline. Participants 4, 6, and 10 were exceptions. Note that each participant solved different tasks. \hl{The median time to solve for the baseline was 936 seconds, while the median time for \tnamei{} was 748 seconds.}

Figure~\ref{fig:timeline} shows a timeline of each participant's session, grouped by task and treatment.
The figure also shows what features of each tool each participant used to solve their tasks over time. Looking at the timeline for the three cases of lower \tnamei{} performance shows that these participants tried multiple packages, searched for examples multiple times and had large amounts of time between executions. In addition, two of these participants (4 and 10) were assigned the same task (task 5).

Participants solving Task 2 and 3 in \tnamei{} were faster than their baseline counterparts\hl{, which can be seen in Figure~}\ref{fig:taskdurations}. Combined with participants 5 (Task 1) and 9 (Task 4), six participants solved their tasks faster using the tool than all participants using \baseline{} for that task. The reverse was only true for two participants, 5 and 9, who completed task 5 faster using the baseline than both participants using the tool.

To explain the performance difference for Task 5, we can look at the task and how participants tackled it. Task 5 has two associated parts, the creation of a graph and the traversal of nodes in the graph. By looking at the commands participants used to search and their transcribed web searches, we can see that three of the four participants who completed this task focused on the graph creation aspect first, neglecting the traversal of the graph. Only participant 9 using \baseline{} immediately searched for graph algorithms, focusing on the depth-first search aspect of the task. Naturally, traversing a graph requires the creation of a graph, leading this participant to find a correct package faster than others. Other participants needed to install and experiment with other packages to come to the conclusion that those packages would not help solve the task as not all packages contain depth-first search functionality. Note that the same participant also performed well when using \tnamei{} in task 4, where they only needed a single package install per task. 

In summary, participants completed tasks in \tnamei{} in most cases faster than they did in \baseline{}.

\subsubsection{How long did participants take to install the first and last package?}

The purpose of comparing the first and last install is to evaluate how long participants spent evaluating packages. The first use of the \texttt{.install} commands marks when participants stop search to try a package, and the last install tells us how long it takes to find a suitable package to complete the task. Where these are the same, it indicates that participants found a suitable package on their first attempt.

\begin{figure}[h]
\begin{center}
    \includegraphics[width=0.95\linewidth]{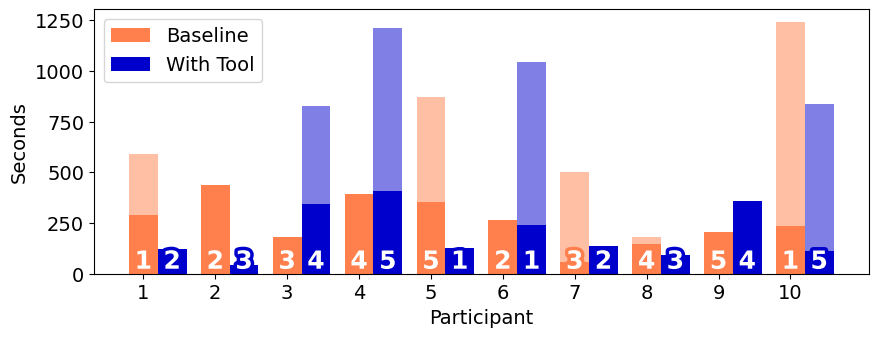}
    \includegraphics[width=0.95\linewidth]{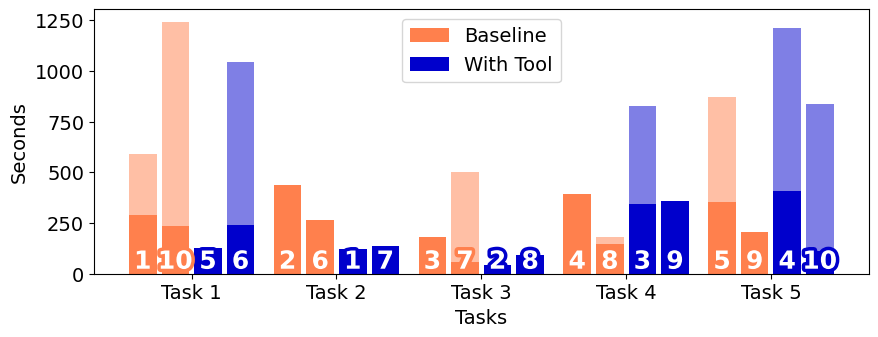}
\caption{Time taken to find the first package (darker), and final package (lighter), \hl{by participant and by task}.}
\label{fig:packagefindtime}
\end{center}
\end{figure}

As described in Section~\ref{sec:user-session}, we record the use of the \texttt{.install} command along with a timestamp. Figure~\ref{fig:packagefindtime} \hl{shows the time to find first and last package, by participant and by task}. Bars of a single colour represent when the first package was also the last package; this is the case for 60\% of participants using \tnamei{} in contrast to 50\% of participants in \baseline{}. We can also see that in 60\% of cases, participants installed the first package earlier using \tnamei{} as opposed to \baseline. Again, in 60\% of cases participants installed the final package earlier in \tnamei{} than they did using \baseline{}. \hl{The median times to find the final package for \baseline{} and \tnamei{} were 250 seconds and 133 seconds respectively.} To sum up, results suggest that, compared to \baseline{}, \tnamei{} enables participants to begin programming faster and to conclude tasks faster.

\vspace{0.5cm}
\fbox{\begin{minipage}{0.85\linewidth}
\textbf{Summary:}~Participants that used \tnamei\ completed tasks without ever leaving the command line interface of the tool. Furthermore, participants using \tnamei\ completed tasks at least as quickly and correctly as they did in the baseline setting where participants had access to web resources such as Google and Stack Overflow.
\end{minipage}}

\subsection{Answering RQ2:~\rqtwo{}}

The aim of RQ2 is to identify developer perceptions about \tnamei{} and its features, such as perceived usefulness which can be important for determining developers' acceptance of new tools~\cite{TAMdavis1989perceived}. The following sections elaborate on the research questions (associated with RQ2) that we posed.


\subsubsection{What was the participant perception of \tnamei{}'s features?}

For this question we investigate what features of \tnamei{} developers found most useful, and respectively, less useful. To measure this, we asked developers a series of questions about \tnamei{}'s features after their session, described in Section~\ref{sec:questions}.

\begin{figure}[h]
\begin{center}
\includegraphics[width=0.9\linewidth]{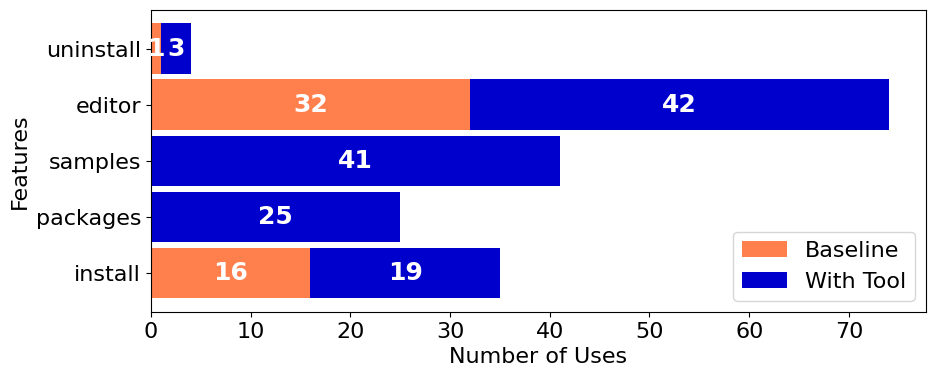}
\caption{Usage of \tnamei{} and \baseline{}'s features}
\label{fig:mostusedfeatures}
\end{center}
\end{figure}

First, we observed how developers used the features of \tnamei{}. Figure~\ref{fig:mostusedfeatures} shows how many times developers used each feature while solving tasks in both the tool and baseline. Recall that the commands \texttt{.samples} and \texttt{.packages} are available only in \tnamei. Overall, the most used feature was the editor mode, used to write code, followed by the feature to retrieve samples. In \tnamei{}, developers looked for packages 25 times, installed a package 19 times and looked for samples 41 times.

\begin{figure}[h!]
\begin{center}
\includegraphics[width=0.95\linewidth]{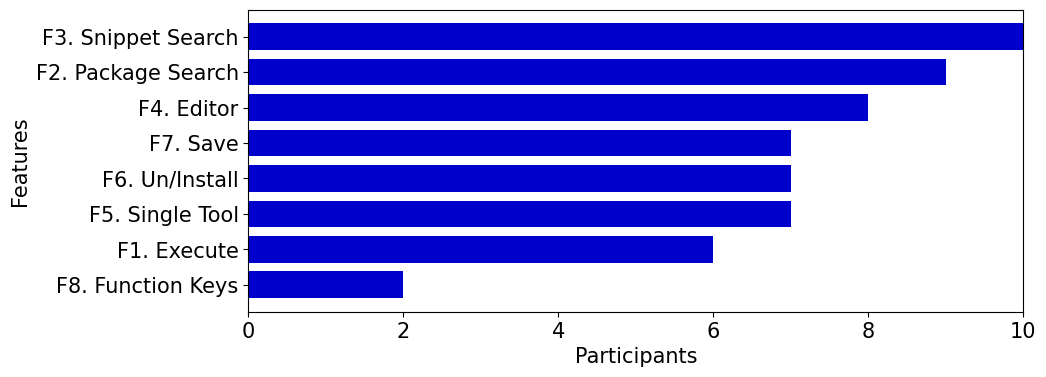}
\caption{What features (Table \ref{table:features}) participants liked.}
\label{fig:liked}
\end{center}
\end{figure}

For question C5, participants were presented with a list of features shown in Table~\ref{table:features}, and asked to select what they liked, where multiple selection was allowed. Figure~\ref{fig:liked} shows how many participants selected each feature. All participants found the code snippet search (F3) to be useful, with the package search (F2) and editor mode (F3) also being well liked. These results show that the \emph{search features of \tnamei} were the highest rated out of all features, and suggest that almost all participants liked the core features of \tnamei{}.

\begin{figure}[h]
\begin{center}
\includegraphics[width=0.95\linewidth]{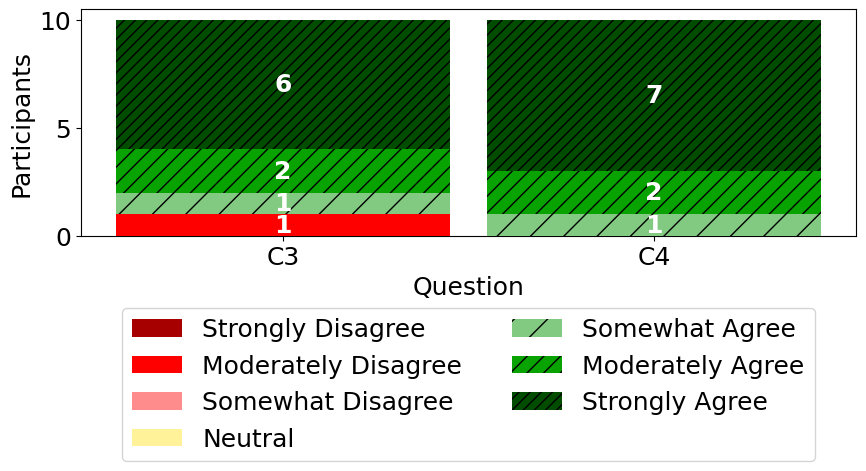}
\caption{Participant ranking of the usefulness of the commands ``packages'' (Question C3) and ``samples'' (C4).}
\label{fig:packsampuseful}
\end{center}
\end{figure}

Similarly, we asked participants to rate the usefulness of specific commands. Figure~\ref{fig:packsampuseful} shows how participants rated the \texttt{.packages} and \texttt{.samples} commands using a seven-rank Likert scale. Overall, participants found both features to be useful. 

We did not ask participants to explain their answers, however, during the session participants indicated issues accessing the function keys from Mac systems, which may explain results for F8 (see Figure~\ref{fig:liked}). During the feedback question, the participant who did not like the packages command indicated that the tool should ``focus on samples''. When asked for feedback at the end of the questionnaire, participants provided suggestions for extending \tnamei{}'s features, such as better package ranking, more interactive UI, and implementing the tool as a plug-in for an existing editor such as Visual Studio Code.

While the editor mode was highly used and well rated, multiple participants indicated that it was limited, lacking familiar controls, and that they would prefer the tool open an external editor instead, with one participant suggesting ``this tool could have a vim mode''.

\subsubsection{What was the general participant perception of \tnamei{}?}

This question investigates the general perceptions of \tname{}. We answered the question in two parts. First, we asked general questions about participant confidence and the helpfulness of \tnamei{} and \baseline{}. Then, we used a popular procedure~\cite{TAMdavis1989perceived,TAMvenkatesh2000theoretical} used to evaluate acceptance of technology to guide our questions to participants.

\begin{figure}[h]
\begin{center}
	\includegraphics[width=0.9\linewidth]{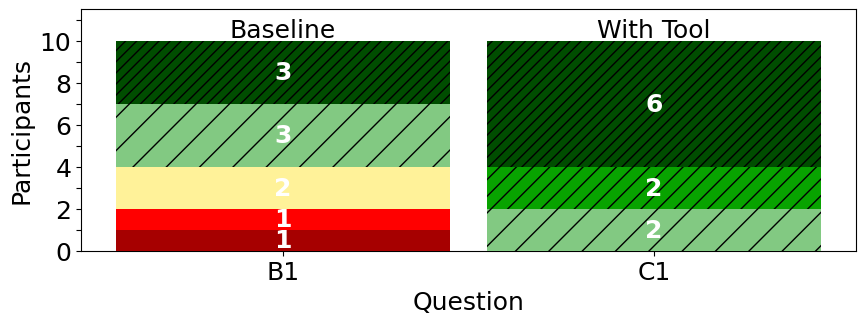}
    \caption{Helpfulness ranking of tool and baseline.}
\label{fig:helpful}
\end{center}
\end{figure}

\begin{figure}[h]
\begin{center}
	\includegraphics[width=0.9\linewidth]{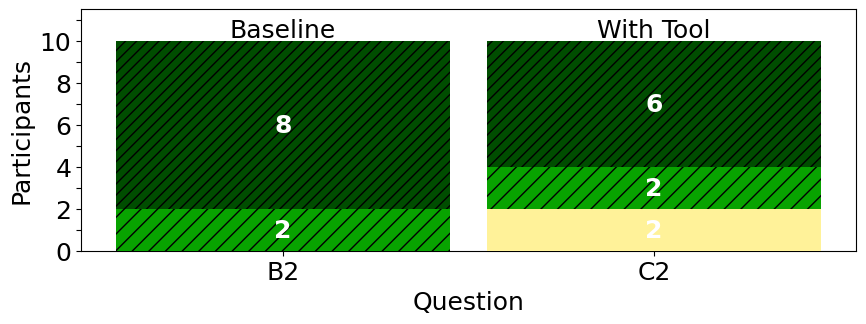}
    \caption{Confidence in task solutions.}
\label{fig:confident}
\end{center}
\end{figure}

\vspace{1em}
\noindent
\textsf{\textbf{General questions:}}~We asked participants to rank how helpful each version of the tool was on a seven-rank Likert scale (B1 \& C1). We also used the same method to measure how confident participants were in their solutions in each tool (B2 \& C2). Figures~\ref{fig:helpful} \& \ref{fig:confident}  show the responses to these questions. Participants found \tnamei{} considerably more helpful than the baseline; all participants agreed that \tnamei{} was helpful, compared to only 60\% for the baseline.

Considering confidence of solutions, results were similar between tools, with most participants confident in their solutions, except for 2 participants answering ``neutral'' for \tnamei{}. There are many factors that may have influenced these results. We expected developers to be less confident doing something unfamiliar, in this case, searching for packages and examples using \tnamei{}. Additionally, all participants completed their tasks successfully in both treatments suggesting that we may have overestimated the time budget. There is also the fact that the design of both tools allows participants to execute code before turn in. These impacts may have leveled the confidence observed within the usage of each tool.

\begin{figure}[h!]
\begin{center}
\includegraphics[width=0.90\linewidth]{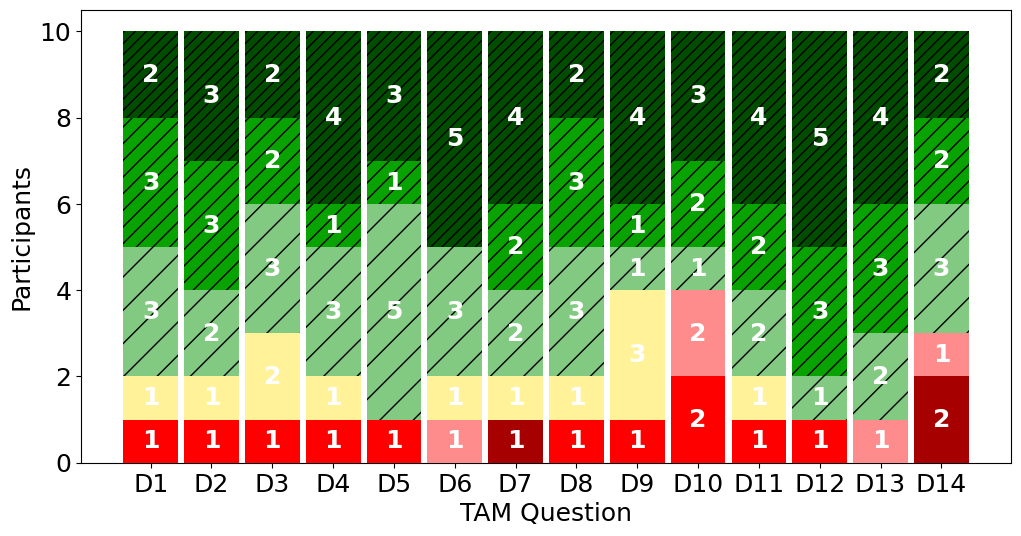}
\caption{Participant response to TAM questions.\protect\footnotemark }
\label{fig:tam}
\end{center}
\end{figure}
\footnotetext{Scale reversed for D14.}

\noindent
\textsf{\textbf{TAM:}}~In addition to assessing developer perception of features, we also surveyed participants using 14 questions adapted from TAM2~\cite{TAMvenkatesh2000theoretical} (Technology Acceptance Model), detailed in Section~\ref{sec:questions}. Figure~\ref{fig:tam} shows the responses for each of these questions.

Overall, developers found \tnamei{} to be useful, and their experiences with the tool were positive. For question D14, most participants agreed they would be able to explain how the tool could be beneficial, One participant responded to all TAM questions negatively; this participant also indicated for question C6 that they found the tool ``a little difficult to navigate''.

\vspace{0.5cm}
\fbox{\begin{minipage}{0.85\linewidth}
\textbf{Summary:} 
Participants' perceptions about \tnamei{} were positive across all evaluated aspects. All 10 participants agreed that \tnamei{} was helpful to them in accomplishing the assigned task (compared to 60\% for \baseline{}). Searching for code snippets and packages were reported as most popular features, and ratings for perceived usefulness, ease of use, and
likely future use (measured by the Technology Acceptance Model) were comprehensively positive.
\end{minipage}}

\section{\tnamei{} Limitations}
\label{sec:Discussion1}

This section discusses the limitations of \tnamei, and how a revised version of the tool, \tnameii{}, addresses those limitations.


The findings from our user study provides evidence that \tnamei{} enables developers to search for and experiment with different packages with reduced context switching. We observed that developers that use \tnamei{} are at least as quick and successful in their coding tasks as developers that followed a ``traditional'' setup with access to search engines, the NPM website, and software documentation from any web resource. We also observed that participants' perceptions are positive across all evaluated aspects, with all participants finding \tnamei{} helpful in accomplishing their tasks.

However, the user studies we conducted revealed that \tnamei{} can be improved in important ways. The following list shows how we addressed user feedback.
\begin{itemize}[leftmargin=*]
    \item Participants expressed that packages highlighted by the existing package search were often not the best choice, as \tnamei{} only uses GitHub stars to rank packages. The results of our user study suggest that when participants cannot find good packages immediately, and instead have to try multiple packages, they spend more time on their tasks. \textbf{Action:~} We address this in \tnameii{} by improving the package search ranking algorithm by incorporating a measure of runnability, (see Section~\ref{sec:improved_search}).
    \item Multiple participants expressed that the UI could be improved, especially the package search UI. Participants mentioned that the table of 25 packages was visually noisy, with too many packages, and that \tnamei{} could be improved by becoming more interactive. In the user study, we observed that most uses of the \CodeIn{.packages} command were followed by the \CodeIn{.install} immediately after, \ie{}, participants installed packages immediately after the search. Participants either copy and pasted the package name as part of the \CodeIn{.install} command, or referred back to the table.  \textbf{Action:~}To address these issues, we replaced the table UI with an interactive, scrollable list of packages that takes up less space, and where users can select a package from the list to install (see Section~\ref{sec:improved_search}).
    \item We observed that most code snippets participants used in both techniques did not run (Section~\ref{sec:rq5}) and participants spent a lot of time making code snippets runnable. \textbf{Action:~}To address this, we implement automated code corrections to reduce the need for developers to make these changes. We also use the error information to rank code snippets (see Section~\ref{sec:code_correction}).
    \item \tnameii{} also adds other, smaller features based on user feedback and real developers' usage patterns, to make the UI more intuitive. We detail these features in Section~\ref{sec:additional_features}.
\end{itemize}

\section{\tnameii{}}
\label{sec:ncq2}


\begin{figure}[t]
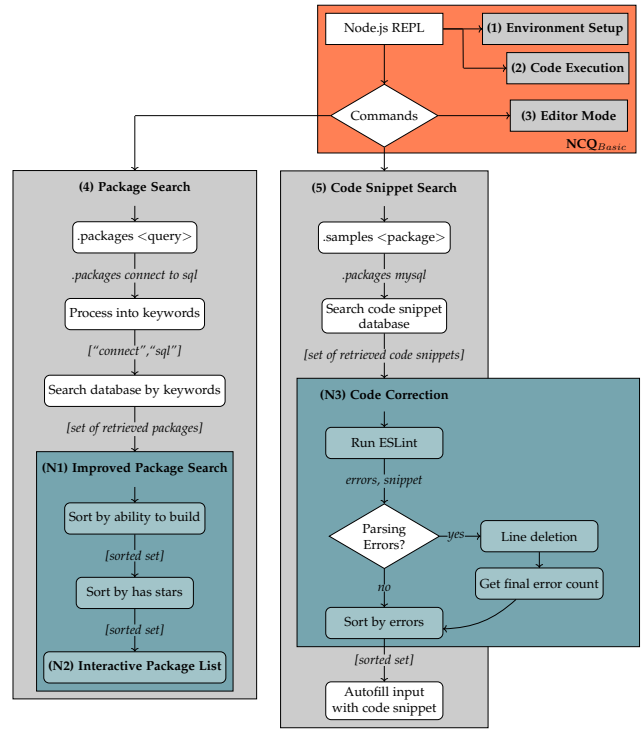

\centering
\includestandalone[width=0.95\linewidth]{figures/ncqfeatures}
\caption{\tnameii{}'s features. The new features of \tnameii{}, in comparison to \tnamei{}, appear in blue.}
\label{fig:pipeline2}
\end{figure}

This section presents \tname{}, a new version of our tool incorporating some of the features requested by the participants from the previous user study.
We focus on two important observations: 1)~package search is an important part of programming in Node.js and 2)~much of the code reuse process depends on modifying and bug-fixing code snippets. 
From those observations, we (1)~improve the package ranking in our search to focus on runnability, not just popularity, (2)~improve the package search with a more interactive UI, (3)~implement measures of code quality and error correction and (4)~make common ways developers use the tool more intuitive. Figure~\ref{fig:pipeline2} provides an overview of the additional features of \tname{}; blue rectangles highlight the new features.
\hl{\tnameii{} retains all the features of \tnamei{}, accessible in the exact same manner, and the additional features do not add any significant performance overhead.}

\subsection{Improved Package Search}
\label{sec:improved_search}
In Section~\ref{sec:Discussion1} we discuss the limitations of \tnamei{}'s package search. We improve upon the star only ranking by implementing a measure of runnability for each package; we used an approach described in previous work~\cite{chinthanet2021makes} to predict the ability to build each package (i.e., install dependencies and prepare working environment) as a proxy for its usability. This approach uses a random forest machine learning model and eight package features from our dataset to predict whether the package is able to build: 
if the package has a license,
if there was a README file,
the number of lines and markdown code blocks in that README,
the number of Node.js code snippets,
if there was a run and install example in the README,
and
the last update.
As Figure~\ref{fig:pipeline2} shows (box ``Sort by ability to build''), the random forest's prediction probability of the ability to build is then used to rank packages in search results. It is worth noting that all packages with no stars are moved to the end of results. We found that 53.7\% of packages in our dataset had no stars, thus this feature enables us to maintain an aspect of (un)popularity in our search.

To address the limitations of the previous package search UI, \tname{} implements an interactive package search. The \CodeIn{.packages} command returns a set of packages; printing a list with all packages on the terminal would make it hard for users to navigate. Our previous solution was to show a subset of packages in a table. \tnameii{} now uses a scrollable list, with terminal height, to show more packages in a smaller space. This interactive design also enables the user to ``select'' a package for installation upon pressing the enter key.

\subsection{Code Correction}
\label{sec:code_correction}
In Section~\ref{sec:Discussion1} we identified that most code snippets needed fixes to run. To address this, we implement automated code corrections based on similar work in NLP2TestableCode~\cite{NLP2TestableCode}. While our previous work focused on \stackoverflow{} snippets and Java, code correction for \javascript{} is a unique problem.

Unlike Java, \javascript{} is not a compiled language, so existing approaches to identify errors like compiler usage cannot be re-implemented here. Instead, existing work in \javascript{} error analysis has used the popular linter ESLint~\cite{campos2019mining}. As a linter, ESLint has code style rules, however, what errors are reported can be customised and it has an automated fix API. We configure ESLint to report three categories of rules: errors for rules that could indicate unrunnable code (for example, the \CodeIn{no-dupe-args} rule), warnings for other erroneous code that does not stop execution but has fixes (e.g. the rule requiring \CodeIn{===} use, which automatically replaces \CodeIn{==} where necessary) and finally a select number of style rules as warnings to enforce a consistent style to all code snippets in \tname{} (indents and semi-colons). We use the warning severity to trigger ESLint's fix API without counting minor issues as errors. We find that with this configuration, 45.2\% of code snippets in our dataset have linting errors, of which 77.2\% are parsing errors which ESLint cannot automatically fix.

In addition to existing ESLint rules, we implement our own. One common error was the use of import/export statements, which the REPL does not support. When running the parser using the \texttt{sourceType} ``script'', to match the REPL functionality, this code triggers a parsing error which cannot be fixed using ESLint. However, if we enable the ``module'' mode, this no longer causes an error. Instead, we implement a custom ESLint rule for this code. We also implement a custom fix that replaces imports with  \CodeIn{require} statements, which most packages also support. With this fix, the error rate reduces to 35.1\% of code snippets. Finally, as most errors were parsing errors with no fixes, we implement a line deletion algorithm  similar to NLP2TestableCode. We comment out lines where parsing errors occur until no more changes can be made or the code snippet has no errors. The result is that 94\% of code snippets have no errors, however 54.2\% of snippets fixed by the line deletion algorithm have all lines commented out. \hl{To address this, we add an additional sorting mechanism that places these 'comment-only' code snippets at the end of the search results.}

Figure~\ref{fig:pipeline2} demonstrates how the error analysis and correction work together. When users search for snippets, each one is run through ESLint's automatic fixes and given an error count. If any of these errors are parsing errors, the line deletion algorithm is run, before a final error count is determined. Finally, the snippets are sorted by error count to prefer error-free snippets.

\subsection{Additional Features}
\label{sec:additional_features}
We add some additional features based on common use cases we saw in the user study. We observed that participants used the editor mode often, and some even pasted snippets directly into the editor. To automate this, opening the editor with empty REPL state now asks the user if they would like to load the previously seen snippet. We also observed that some code snippets with constant variables, often used when importing a package, caused issues for participants when trying to rerun snippets. This would cause cases where code ran once but not again. To help with this, we renamed the default \CodeIn{.clear} command to \CodeIn{.reset} (to differentiate it from ``clearing the input'') and included an additional error message for this case reminding them they could reset the state of the REPL.


\section{\tnameii{} Evaluation}
\label{sec:eval2}


To evaluate \tnameii{}, we conducted a user study involving 10 Node.js developers. This time, we compared use of \tnameii{} to the current state of practice, a traditional editor (Visual Studio Code) and online search, instead of an artificial baseline. Section~\ref{sec:overview} describes the baselines for each evaluation. Similar to the previous evaluation, participants were asked to complete two coding tasks with and without \tnameii{} and, after each task, they were asked to answer questions about their experience.

We ask the following research questions about participant performance and perceptions, similar to RQ1 and RQ2, as well as an additional question to analyse \tnameii{}'s new code correction feature:

\vspace{1ex}
\noindent
\textbf{RQ3.} \rqthree{}

\noindent
\textbf{RQ4.} \rqfour{}
\vspace{1ex}

\noindent
\textbf{RQ5.} \rqfive{}
\vspace{1ex}

\subsection{Experimental Design}

We maintain the same design as the first study, except where necessary changes were made due to the use of a traditional editor. Participants were given the same tasks, with the same assignment and most post-task questions remain unchanged. This section describes the differences:

\subsubsection{Baseline}
\label{sec:baseline2}

We compare \tnameii{} to the baseline of a traditional editor and internet access. The aim is to measure how \tnameii{} compares to a more realistic baseline consisting of developers programming in a real editor and using online search to find packages and examples. We selected Visual Studio Code, the most popular editor among Node.js Developers, considering the responses we obtained in our survey (Section~\ref{sec:survey}). We setup VSCode as a clean installation with no extensions, open to a new Node.js project and empty ``index.js'', with the integrated terminal open for executing Node.js code.

\subsubsection{Participants}

Due to difficulties finding participants for the first user study, we decided on a more rigorous recruitment process, using the Prolific platform to recruit 10 participants. We verified that participants had basic Node.js skills using a screening survey where participants were asked programming questions, which we describe more in-depth in a workshop paper~\cite{prolificUserStudy}. Participants had an average of 3 years experience. The recruited participants are a subset of the same 55 developers who answered our initial survey questions in Section~\ref{sec:survey}.

\subsubsection{Questions}

Participants were asked to answer questions after each task. This time, we already had their demographic details from recruitment, so omitted section~A. We also omitted question B1 (helpfulness of the baseline tool) as participants now use VSCode for their first task. Questions can be seen in Table~\ref{table:questionsABC2}.

\begin{table}[h]
    \centering
    \caption{Questions asked (B) after task~1, and (C) task~2.
    }
  \rowcolors{3}{white}{LightGray}
    \begin{tabularx}{\linewidth}{lX}
    \toprule
    \hiderowcolors 
    \showrowcolors
    \hiderowcolors 
    \showrowcolors
    C1 & Overall, did you consider the tool helpful to accomplish the assigned task? (1 (Strongly Disagree) - 7 (Strongly Agree)) \\
    B1/C2 &  Grade your confidence, in a 1-7 scale, that your solution is correct according to the problem specification? \\
    C3 & Did you consider the command ``packages'' useful? \\
    C4 & Did you consider the command ``samples'' useful? \\
    C5 & What features did you like about the tool? \\
    C6 & What you did not like about the tool? \\
    \bottomrule
    \end{tabularx}
    \label{table:questionsABC2}
\end{table}

At the end of the session participants were asked to answer the same TAM questions as in the first evaluation.

\subsection{Answering RQ3:~\rqthree{}}

The goal of RQ3 is to identify the helpfulness of \tnameii{}'s features and how those features are used by developers. As in RQ1, we hypothesise that integrating package search within a programming environment reduces context switching, improving developer's performance. To answer this question, we compare how participants performed searching for packages using \tname{} and using the internet when programming in VSCode.

\subsubsection{Was there an impact on time to complete each task?}

\begin{figure}[h!]
\begin{center}
\includegraphics[width=0.95\linewidth]{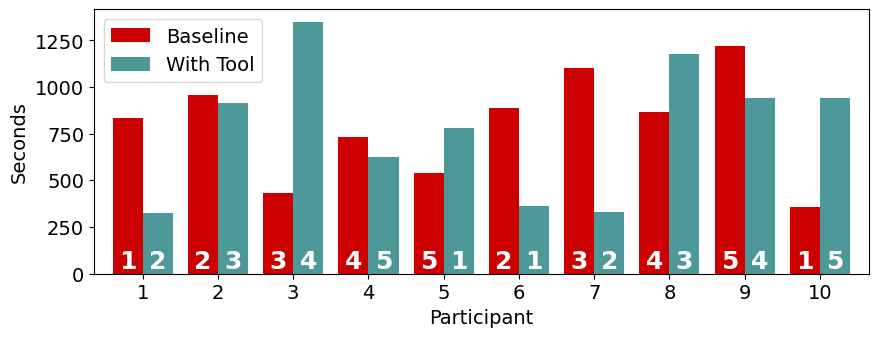}
\includegraphics[width=0.95\linewidth]{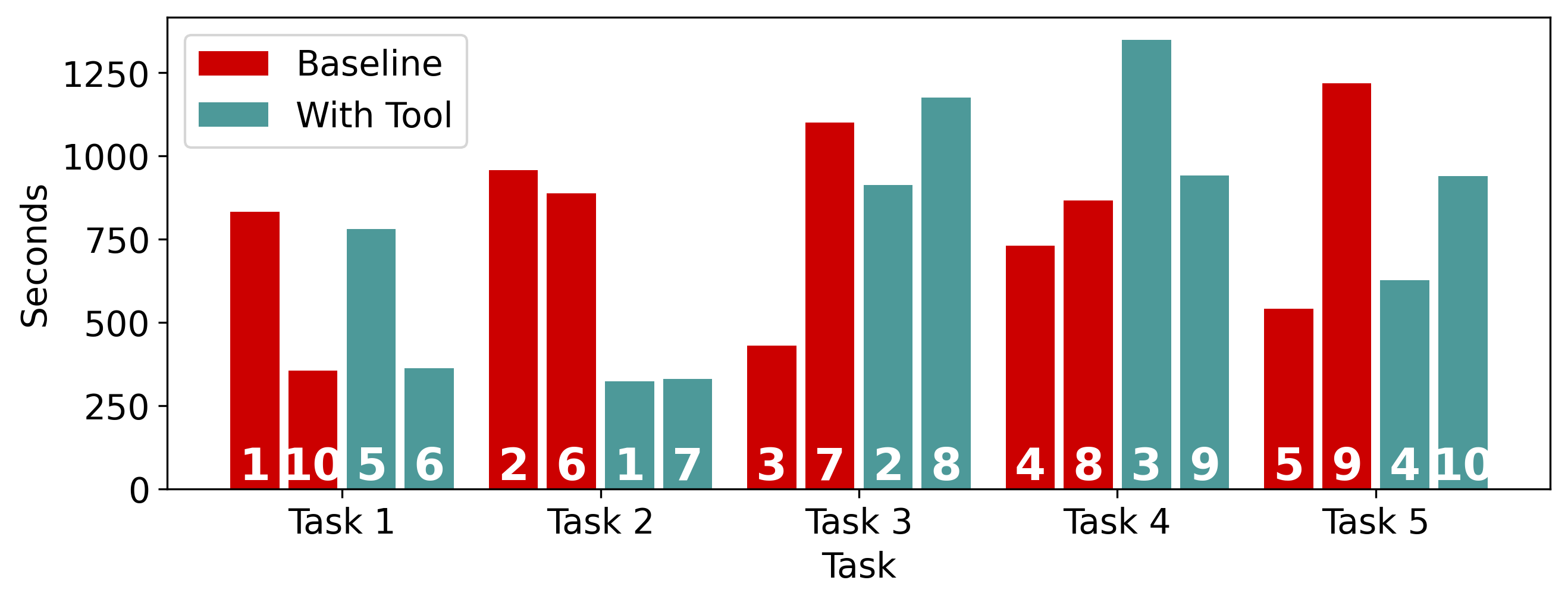}
\caption{\hl{Duration of tasks by participant and by task. Bars are labelled by task ID in the first figure, and by participant ID in the second figure.}}
\label{fig:taskdurations2}
\end{center}
\end{figure}

Figure~\ref{fig:taskdurations2} shows the task durations for both \tnameii{} and the baseline. \hl{Again, all participants solved their tasks.} Participants completed tasks in \tnameii{} faster than in a traditional editor in 6 out of 10 cases. \hl{The median time to solve for the baseline was 849 seconds, while the median time for \tnameii{} was 847 seconds.}

The timeline in Figure~\ref{fig:timeline} details each session, with participants 1-10 belonging to the first user study and participants 11-20 belonging to this study. The icons remain the same as in the first study. Participants using VSCode + Web (note the `search' marker in the timeline figure) follow links to evaluate the results of their search as opposed to cycling through snippets. In \tname{} versions participants can press enter to execute code or use the editor mode, which we label as `submit' or `editor' in the timeline. In VSCode, we instead transcribed when participants used the terminal to `execute' their code, and when they pasted code snippets from online into their open file. However, not all participants pasted code snippets, instead some manually rewrote code while making changes (e.g. 13, 15, 16 and 20).

The three fastest sessions using \tnameii{} (11, 16, 17) involve only one install and snippet search, conversely, the longest sessions (13, 18, 20) have many uses of the snippet search. Again, as in the previous user study, we observe that the number of times participants search may negatively impact their task duration.

\subsubsection{How long did participants take to install the first and last package?}

\begin{figure}[h!]
\begin{center}
    \includegraphics[width=0.95\linewidth]{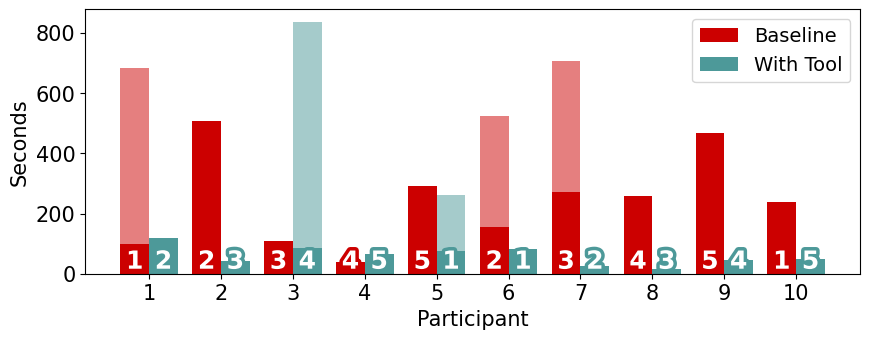}
    \includegraphics[width=0.95\linewidth]{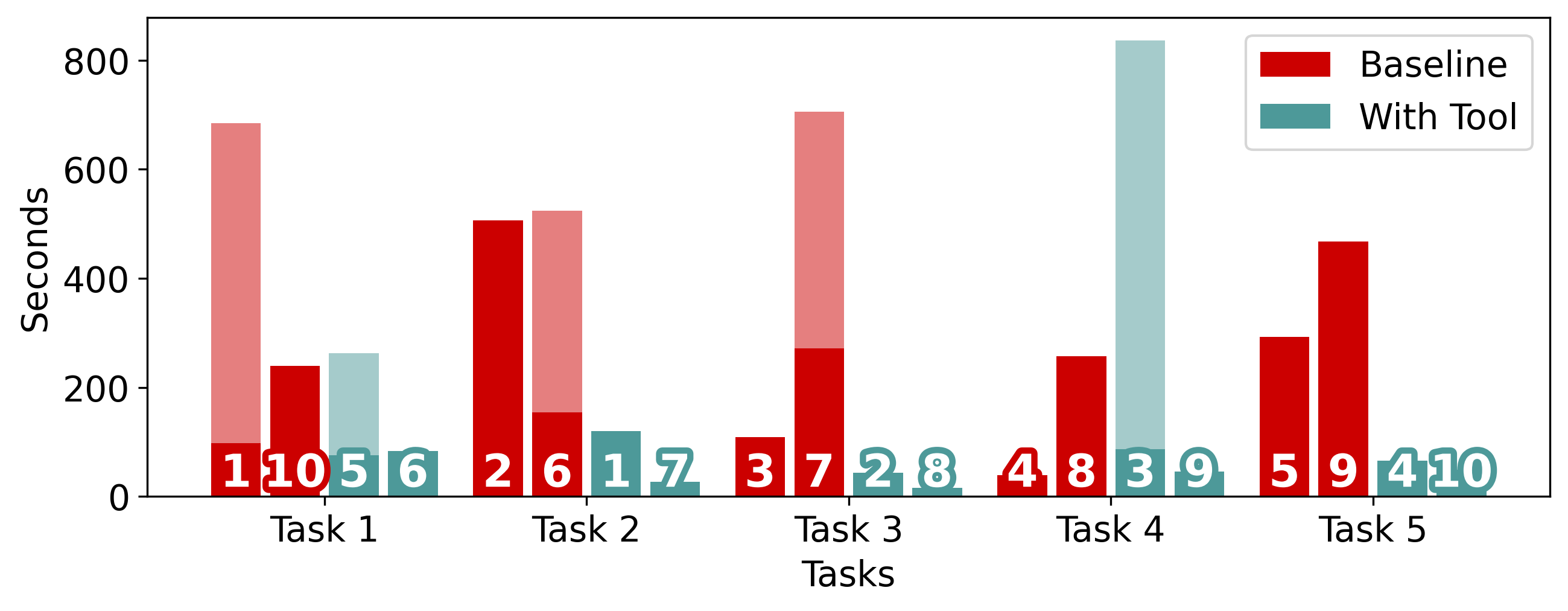}
\caption{Time taken to install the first package (darker), and final package (lighter), \hl{by participant and by task}.}
\label{fig:packagefindtime2}
\end{center}
\end{figure}

Figure~\ref{fig:packagefindtime2} shows the times the first and last packages were installed. This information reveals how long participants take in the ``package search'' phase of their tasks. We observed that the final installed package was the package used to complete the task in all cases. Participants' first and last packages were the same (only 1 package installed) in 8 cases for \tnameii{} and 6 using VSCode. In 7 cases participants found their last package faster using \tnameii{} than VSCode. \hl{The median times to find the final package for the baseline and \tnameii{} were 380 seconds and 57 seconds respectively. These results highlight the efficiency of finding packages in \tnameii{}.}

\vspace{0.5cm}
\fbox{\begin{minipage}{0.85\linewidth}
\textbf{Summary:} 
Participants completing tasks using \tnameii{} performed better than when using VSCode. They were able to solve tasks and find packages faster, in most cases finding a suitable package immediately.
\end{minipage}}

\subsection{Answering RQ4:~\rqfour{}}

The goal of RQ4 is to identify developer perceptions of the tools, such as helpfulness, and confidence in their solutions. We ask the following questions:

\subsubsection{What was the participant perception about the features that \tname{} offers?}

The goal of this question is to understand participant perceptions of each of \tnameii{}'s features and to identify which features were useful or not.

\begin{figure}[h]
\begin{center}
\includegraphics[width=0.9\linewidth]{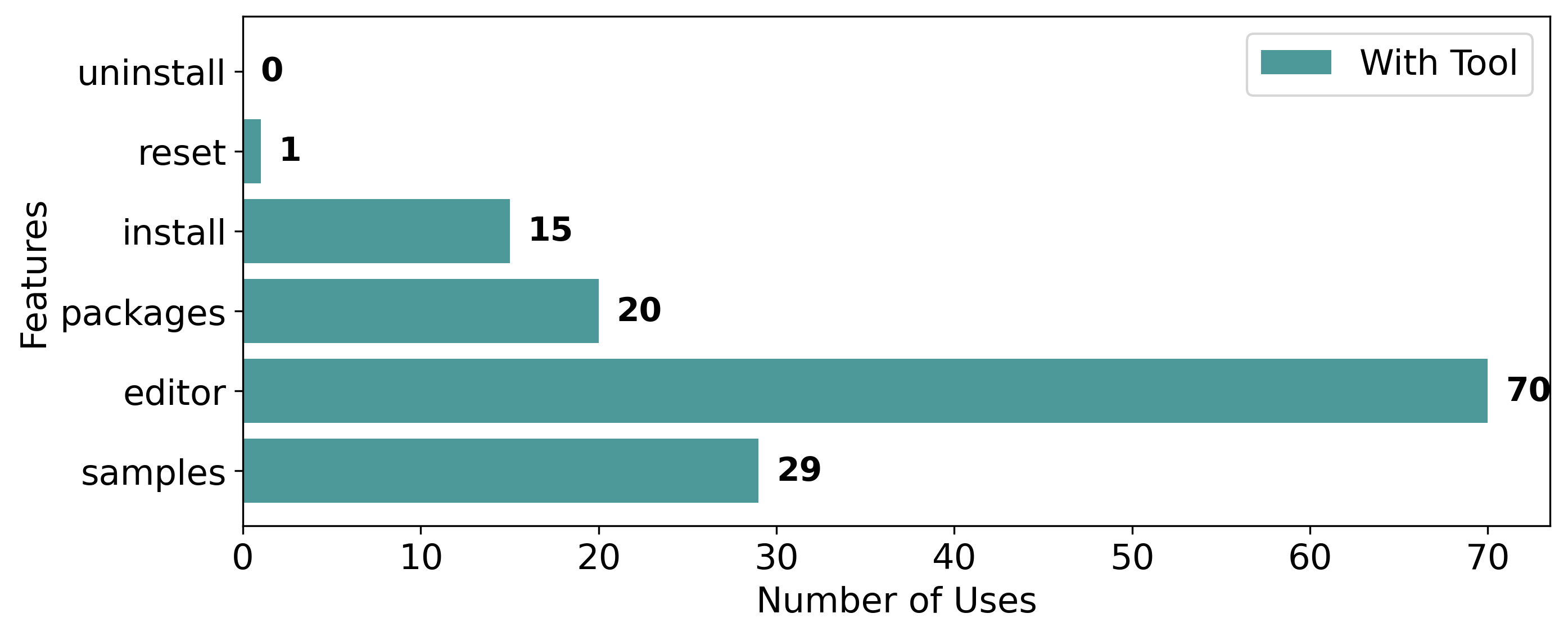}
\caption{How many times \tnameii{}'s features were used.}
\label{fig:mostusedfeatures2}
\end{center}
\end{figure}

Figure~\ref{fig:mostusedfeatures2} shows the features 
that participants used the most. As in the first study, the \emph{editor} mode and the search for \emph{samples} remained the most popular features. Relative to the first study, there was an increase in the total number of uses of the editor mode, but a reduction in the use of the  commands \CodeIn{.packages} and \CodeIn{.samples}. In the first case, the reduction may be attributed to the changes to the editor mode whereas, in the second case, the reduction may be justified by the fact that more participants solved tasks faster and with the first package, reducing the need to look at more packages and their examples.

\begin{figure}[h!]
\begin{center}
\includegraphics[width=0.95\linewidth]{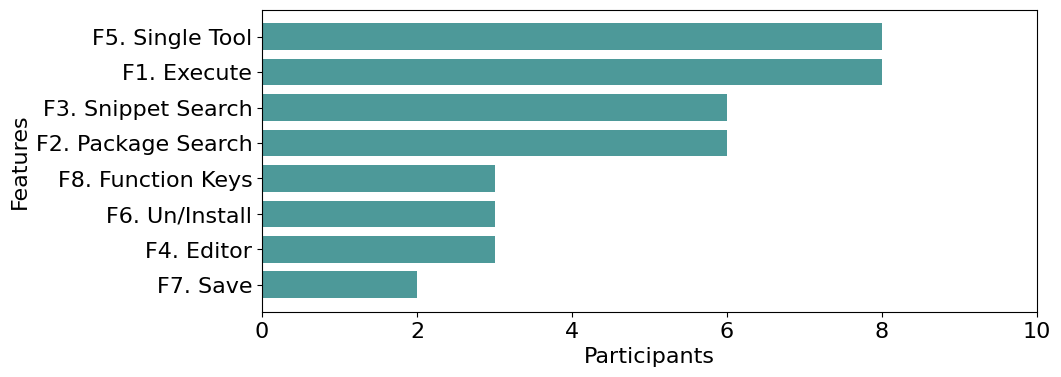}
\caption{Number of participants that liked each feature of \tname{}. Features (F) are listed in Table \ref{table:features}.}
\label{fig:liked2}
\end{center}
\end{figure}

Recall that Table~\ref{table:features} shows the list of \tname's features. According to question C5 (see Table~\ref{table:questionsABC2}), the features that participants liked the most were the ``ability to execute code in the REPL''~(F1) and ``the combination of code search and execution in a single tool''~(F2). The feature participants liked the least was F7, ``the ability to save and resume''. This is a shift from the first study, where snippet, package and editor mode were the most highly rated. In fact, the editor mode was only selected by three participants despite being used frequently. In general, most participants liked the core features of \tnameii{}: the package (F2) and code snippet search (F3), REPL execution (F1) and combination of search and execution environment (F5).

\begin{figure}[h]
\begin{center}
\includegraphics[width=0.9\linewidth]{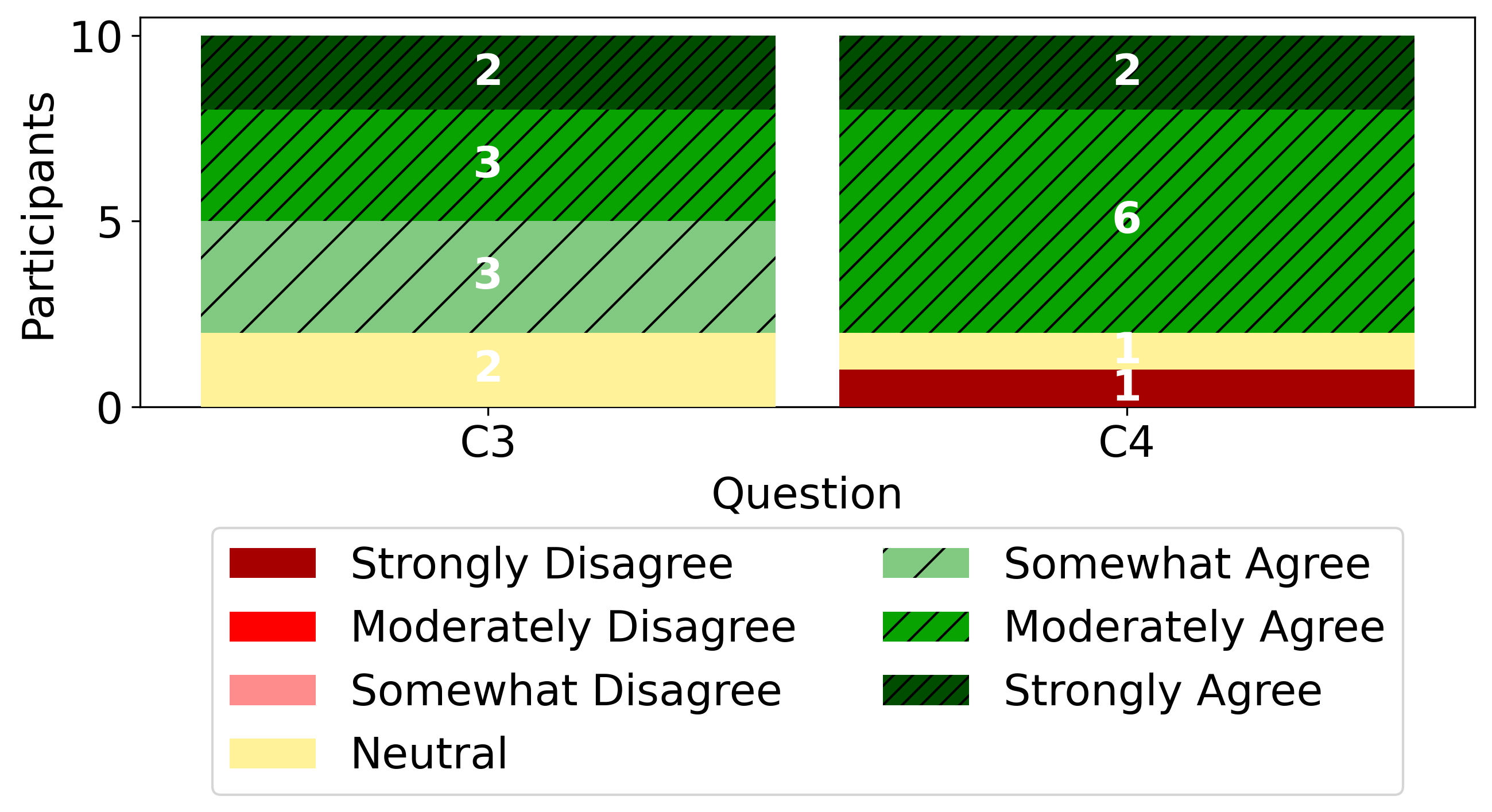}
\caption{Participant ranking of the usefulness of the commands ``packages'' (question C3) and ``samples'' (C4).}
\label{fig:packsampuseful2}
\end{center}
\end{figure}

Figure~\ref{fig:packsampuseful2} shows how participants ranked the usefulness of commands \texttt{.packages} and \texttt{.samples} using the same seven rank Likert scale from the previous user study. Eight participants agreed that both features were useful, with only one strongly disagreeing with the samples feature (this participant then answered for question C6 that they thought showing only code snippets and not their surrounding documentation which ``exists to guide the user'' was limiting). Despite this feedback, all participants were able to successfully complete their tasks in \tname{}, and the results of our user study suggests they performed similarly or better than when they had access to this documentation in the baseline. While we cannot conclude that a lack of documentation has a positive impact on performance, in fact, pairing documentation and code snippets may have improved performance further, we can argue that participants were not limited greatly because of it. In general, there was a slight decrease in agreement between studies, but the result is still very positive.

\subsubsection{What was the general participant perception of NCQ?}

\begin{figure}[h]
\begin{center}
	\includegraphics[width=0.42\linewidth]{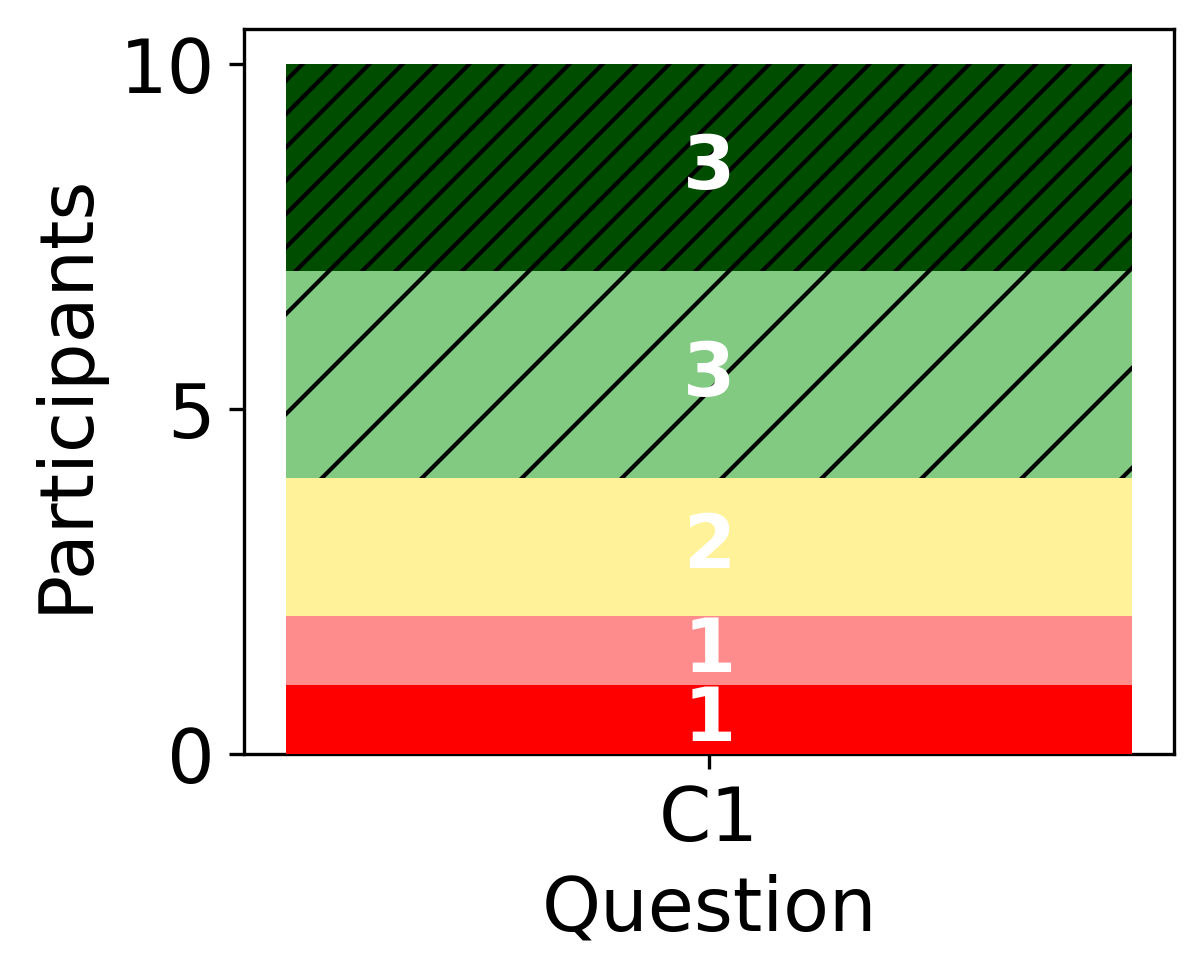}
    \caption{Participant rankings for helpfulness of \tnameii{} (C1)}
\label{fig:helpful2}
\end{center}
\end{figure}

We asked participants to rank the helpfulness of \tnameii{}; Figure~\ref{fig:helpful2} shows the responses. Six out of ten participants agreed that \tname{} was helpful in solving their tasks, and only two participants disagreed. Agreement fell compared to the previous study, however. We did not ask participants to rank the helpfulness of the baseline, VSCode, in this task, but we conjecture that the comparison to a traditional editor impacted participant perception of \tname{}.

\begin{figure}[h]
\begin{center}
	\includegraphics[width=0.9\linewidth]{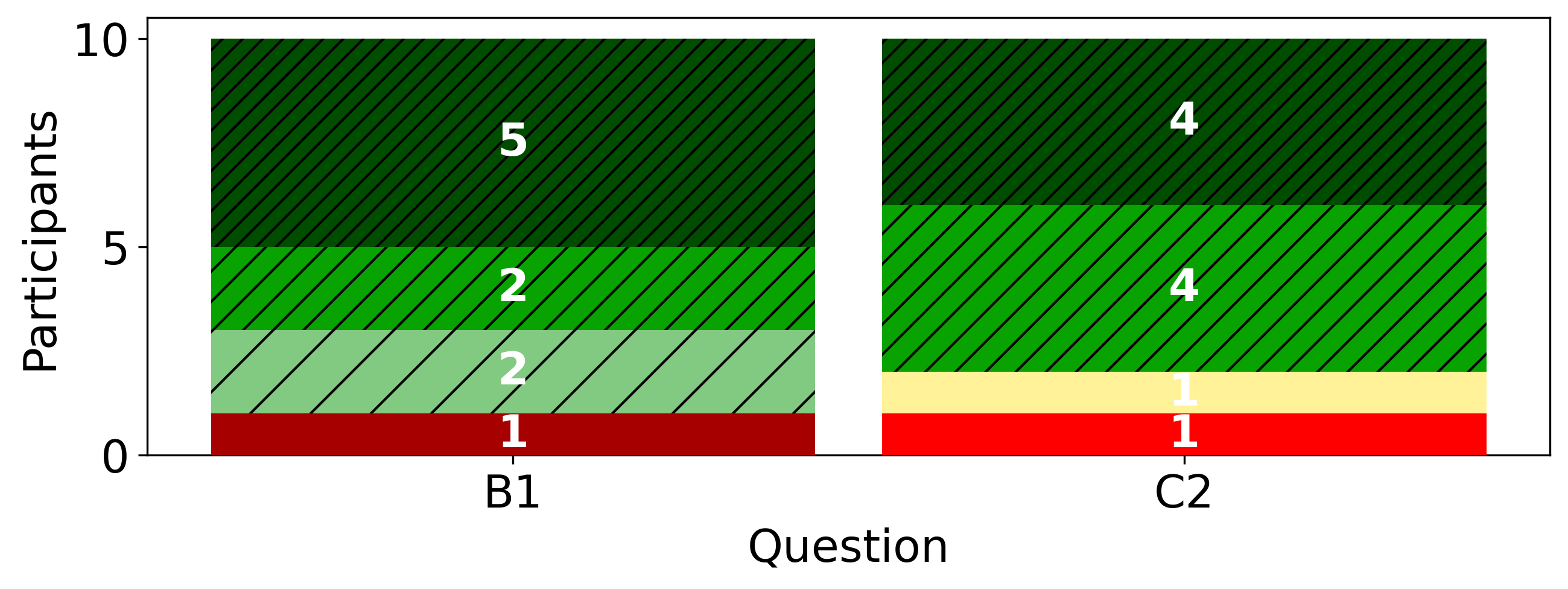}
    \caption{Participant confidence ratings (B1 \& C2).}
\label{fig:confident2}
\end{center}
\end{figure}

Figure~\ref{fig:confident2} shows the confidence (1-7 scale) of participants in their solutions for each technique. We observe that confidence is similar between techniques; nine participants scored their confidence highly for the baseline (score of 5 or above), compared to eight for \tname{}. Individually, four participants had their confidence increase or stay the same, with the largest increase from a score of 1 to 7. For four other participants their confidence only fell by one point. This decrease in confidence is also consistent with the first study. Overall, participant confidence in solutions did not suffer considerably despite the use of a new tool when compared to a more familiar baseline.

\begin{figure}[h]
\begin{center}
\includegraphics[width=0.9\linewidth]{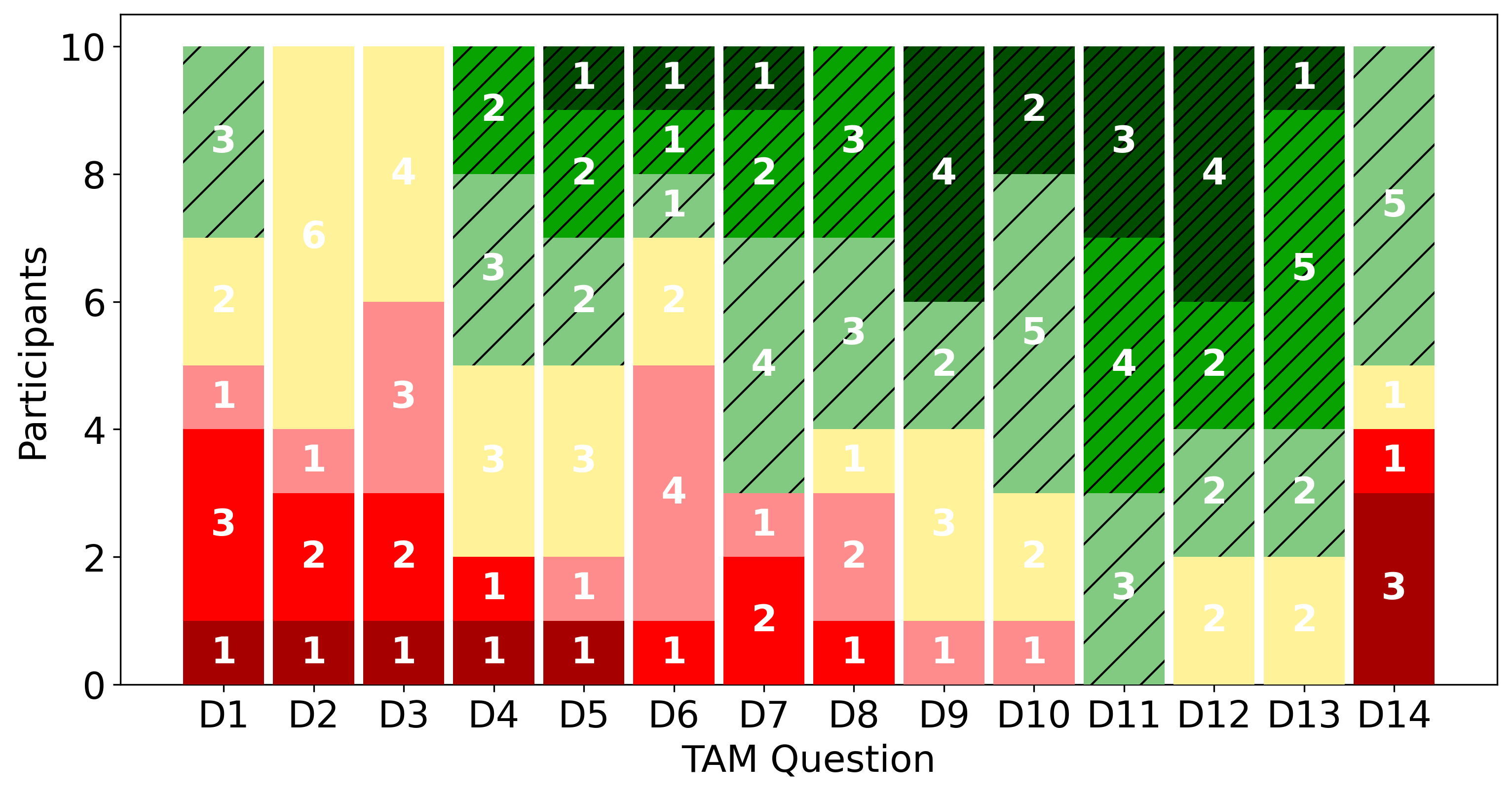}
\caption{Participant response to TAM questions.\protect\footnotemark }
\label{fig:tam2}
\end{center}
\end{figure}
\footnotetext{Scale reversed for D14.}

Participants were asked the same TAM questions as the first study. Results were not as overwhelmingly positive as the first study, however, responses were still more positive than negative, as can be seen in Figure~\ref{fig:tam2}. All participants agreed with D11 (no difficulty telling others about the results of using the tool), and most participants agreed with D7 (easy to use), D10 (high quality of output), D12 (could communicate to others the consequences of using the tool) and D13 (results of using the system are apparent). However, most participants disagreed with D3 (using the tool would enhance my effectiveness in my job), and six participants agreed with D14, that they would have difficulty explaining why the system may or may not be beneficial. 

\vspace{0.5cm}
\fbox{\begin{minipage}{0.85\linewidth}
\textbf{Summary:} 
Participant perception of \tnameii{} was generally positive; most participants were confident in their solutions and found the tool helpful, despite the comparison to a more familiar baseline, VSCode. Using the Technology Acceptance Model to measure, most participants found \tname{} easy to use, with a high quality of output.
\end{minipage}}

\subsection{Answering RQ5: \rqfive{}}
\label{sec:rq5}

The goal of this research question is to evaluate the impact of \tname{}'s code correction features on code snippet quality. First, we observe that 33\% (27/81) of code snippets retrieved by participants using \tnameii{} had errors fixed. To measure the effect of this, we compare the quality of code snippets participants interacted with across all sessions, including the previous user study. We reason that \tnameii{} should improve the quality of code snippets through its code correction features.

For the purpose of this analysis, we define:
\begin{itemize}[leftmargin=*]
    \item ``seen'' as the set of all code snippets participants encountered during their task solving, which they evaluate to select snippets to use. For \tname{}, we counted all code snippets that participants saw using the \CodeIn{.samples} command, which were automatically logged. However, for online search, we manually logged from video all snippets that were visible in the web browser. Partial snippets (those that the participant did not scroll down far enough on a web page to see entirely) are excluded as we do not have the entire snippet to evaluate. Likewise, we argue that participants could not have evaluated these snippets for use without seeing them in full.
    \item ``attempted use'' as the subset of ``seen'' that participants attempted the code reuse process with; that is, participants copied these snippets into their editor, before making changes and/or running them.
    \item ``successful use'' as a subset of ``attempted use'', where any part of a code snippet was successfully used by a participant to complete their assigned task. 
\end{itemize}

To evaluate the quality of code snippets, we look at both linting and runtime errors. We use the same ESLint configuration used in \tnameii{} to detect linting errors. To measure the presence of runtime errors, for each code snippet, we created a Node.js project, installed any needed packages and ran the code snippet in a file. While existing work has shown that most code snippets in documentation do not execute~\cite{chinthanet2021makes}, instead many are missing parts and may not be intended to be fully working examples, we still consider this measure to be useful for comparison purposes. Developers reusing example code must still transform unrunnable code into runnable code, and runtime errors may represent work needed. The goal of automated error correction is to reduce this work, making runtime errors a useful measure.

\begin{figure}[h]
    \centering
    \includegraphics[width=\linewidth]{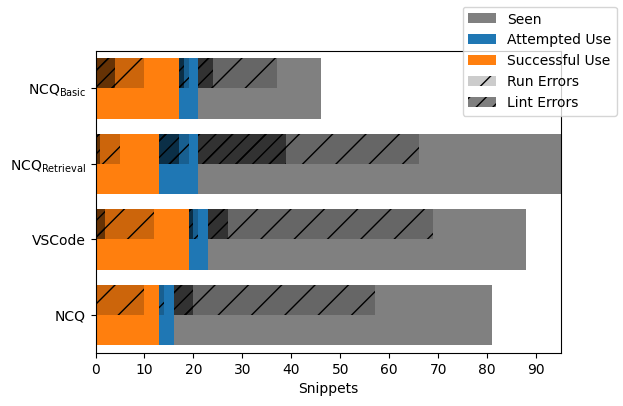}
    \caption{Code snippet breakdown for each technique.}
    \label{fig:snippet_errors}
\end{figure}

Figure~\ref{fig:snippet_errors} compares the two baselines and two \tname{} variations from both evaluations. We found that in the first user study, 60\% of seen and 55\% of used code snippets did not run, and that 22\% of seen snippets and 24\% of used snippets have linting errors. For the \tnameii{} user study, the runtime error rate remains similar for seen and used at 66\% and 64\% each, however, we see a large reduction in linting errors with the code correction features of \tnameii{}. While the number of seen snippet linting errors in the baseline remains similar at 12\%, code snippets found using \tnameii{} had an error rate of 8\%. Even more, none of the snippets participants used had any linting errors. We also see a 5\% increase in the proportion of used snippets that were used successfully, which is not seen in the first user study.

Additional observations from this analysis can be made. In the first study, using \tnamei{}'s search features, participants were able to see more code snippets more quickly than in the baseline, despite attempted use rate remaining the same. However, in \tname{}, we see that participants saw less snippets, and also attempted to use less snippets, than in the baseline, following the trend of participants being more efficient while using \tnameii{}.

For all techniques, we see that there are more code snippets with runtime errors than there are with linting errors. This is a strong indication that there is room for improving the error detection in \tnameii{}, and in improving static analysis in JavaScript in general.

\vspace{0.5cm}
\fbox{\begin{minipage}{0.85\linewidth}
\textbf{Summary:} 
\tnameii{}'s code correction features were employed on 33\% of code snippets, and reduced the number of snippets with errors seen and used by developers; all code snippets participants used in \tnameii{} had no linting errors. However, we saw no reduction in runtime errors, indicating there is room for better static analysis of JavaScript code for error detection and thus error correction.
\end{minipage}}

\section{Discussion}

Developers increasingly rely on gluing existing code together as opposed to developing new code~\cite{abdalkareem2017developers}. Node.js developers in particular have access to a rich ecosystem of 1M+ software packages through NPM. However, navigating this wealth of potential code solutions is a challenge. One user study participant summed up their experience with: ``Sometimes there isn't enough information on each package, and it is very tiring to install/uninstall everything before having a good idea of what it does''. Searching and installing packages is a time sink, as described by another participant: ``It's not unusual to spend more than a [sic] hour looking for a package for simple job''. Yet, installing packages and running corresponding code snippets are crucial activities in deciding whether to use a package. Alleviating these challenges is the goal of \tnameii{}, as summarised by a participant: ``Finding a suitable package can be difficult given the volume of packages available. It can become tiresome installing packages and trying out code, constantly referring back to documentation or the dedicated package page on npmjs.com. The .samples feature of \tname{} looks to reduce the time taken to test packages, to determine if the package provides benefit, and is suitable for a given project.'' When using \tnameii{}, developers do not need to context switch between editor and search engine; instead they can explore packages, install and uninstall them, and run code snippets from within the same command line interface.

\tnameii{} supports developers' desire to try out code snippets quickly and experiment with different potential solutions when embarking on a new task. Even for trivial tasks, developers commonly use third-party packages~\cite{abdalkareem2017developers}, and \tnameii{} supports this workflow. We observed that the typical usage pattern participants followed in \tnameii{} was to search for packages, then search for samples within those packages. The example in Section~\ref{sec:automatedexample} describes such a workflow based on a real user study session, and we observed similar workflow for online search; most participants started by looking for packages and package documentation on NPM or GitHub, with only a few accessing tutorial sites. This reveals the need for Node.js specific development tools to fit the workflow of Node.js developers.

For \tnameii{}, we improved upon \tnamei{} based on the feedback obtained from users of the tool. The results of \tnameii{}'s evaluation indicate that the tool enables developers to be more efficient finding packages than both \tnamei{} and a traditional editor combined with internet access. For example, the average task completion duration reduced from 1,045 seconds for \tnamei{}, to 774 for \tnameii{}. However, despite this performance improvement, we see that developer perceptions of \tnameii{} were harsher. There are multiple factors that may explain this. For the first study, most participants were students and we did not verify Node.js skill; in contrast, for the second study, we made the additional effort to recruit participants with a minimum level of Node.js skill while trying to recruit non-student programmers. This difference in experience between the two groups may impact perceptions of the tools. Also, the different baselines may have impacted how developers perceived each tool. In the \tnameii{} evaluation we directly contrast the tool to a more familiar baseline; participants use \tnameii{} directly after using VSCode. However, in the \tnamei{} evaluation participants use a baseline that is simply \tnamei{} with reduced features; there may be a perceived `improvement' between the two tasks, and participants were not faced with a direct comparison between \tnamei{} and tools they use regularly to program. Despite all of this, results did remain similar, and in most cases still positive.

Our findings provide evidence that \tnameii{} still performs well even when directly compared to a familiar, professionally developed baseline such as VSCode. Our evaluation confirms that \tnameii{} was well received by developers because it reduces the need for context switching typically associated with reusing third-party code.

\section{Related Work}

This section discusses related work in the areas of code reuse, code search, library selection and code executability, as \tname{} makes contributions to all of these aspects combined into a single tool.

\subsection{Empirical Studies on Code Reuse}

Various studies have been conducted to understand developers' perceptions on the advantages and disadvantages of code reuse~\cite{basili1996reuse, mockus2007large, lim1994effects, mohagheghi2004empirical, sojer2010code,li2016relationship}. For example, Sojer and Henkel~\cite{sojer2010code} surveyed 686 open source developers on their reuse practices. They found that experienced developers adopt reuse practices more regularly than non-experience developers and 30\% of the functionality of open source software projects are based on components from different projects. Overall, developers perceive code reuse as an important practice, especially when starting new projects. \tname\ builds on the observation made in previous studies that there is a vast amount of code examples on the web today and developers do try using them~\cite{NLP2TestableCode}. \tname{} provides a platform for code reuse in Node.js where developers can experiment with different examples. It is worth noting that, although \tname{} uses README files from NPM to obtain code snippets, additional sources would be easy to integrate in future work. For example, other work has focused on mining usage examples from Q\&A forums like Stack Overflow~\cite{NLP2TestableCode}, or where examples in documentation are insufficient, mining API usage examples from source code~\cite{mcmillan2010recommending, moreno2015can, SAIED2018164}.

\subsection{Code Search}

Code search has been extensively investigated recently~\cite{Campbell2017NLP2CodeCS,NLP2TestableCode,ponzanelli2014prompter,zhang2016bing,brandt2010example,ponzanelli2013seahawk, gu2018deep, inIDECodeGen}. Brandt~\etal{}~\cite{brandt2010example} proposed Blueprint, an in-editor tool that looks for code examples on the web. Experiments found that it enabled developers to write higher quality code and that search with Blueprint was significantly faster compared to regular web search. Similar to Blueprint, Seahawk~\cite{ponzanelli2013seahawk} finds Stack Overflow code snippets in the editor, using the existing code context to automatically formulate search queries. Prompter~\cite{ponzanelli2014prompter} automatically notifies the developer when it finds snippets on Stack Overflow that are similar to parts of the code under development. Campbell and Treude proposed NLP2Code~\cite{Campbell2017NLP2CodeCS}, an approach where developers use natural language queries to find code snippets. NLP2Code is integrated in the development environment to reduce context-switching, provides task suggestions and works entirely in the editor, with no additional views or windows. Reid~\etal{}~\cite{NLP2TestableCode} proposed NLP2TestableCode, an improvement over NLP2Code that checks whether code snippets contain errors (\eg{}, type errors and basic integration errors) and, if so, attempts error repair before recommending the snippets to users. Results indicate that the tool can find compilable code snippets in 82.9\% of cases and it can find ``testable'' code snippets (able to be converted into a function with input/output for testing) in 42.9\% of cases. \tname{} implements a similar code correction functionality, however, our code correction is limited by the lack of comprehensive error reporting in Node.js.

Some recent approaches also utilise machine learning; Gu et al.~\cite{gu2018deep} proposed DeepCS, an approach that uses deep neural networks for code search. DeepCS uses a neural network model, dubbed CODEnn, embedding code snippets and natural language descriptions into a high-dimensional vector space. The intuition is that CODEnn incorporates semantic information to the search in contrast to typical Information Retrieval approaches, which only use text similarity for search. Preliminary results showed that DeepCS outperforms alternative code search techniques. Kanade~\etal{}~\cite{DBLP:conf/icml/KanadeMBS20} proposed and evaluated CuBERT, a code-understanding BERT model, trained with a deduplicated corpus of 7.4M Python files from GitHub. CuBERT was compared against state-of-the-art alternatives to solve five classification tasks and one code repair task, showing superior performance, even with shorter training and with fewer labeled examples. It remains to evaluate CuBERT's performance for code search. Xu~\etal{}~\cite{inIDECodeGen} proposed NL2Code, an in-editor tool that combines code retrieval and code generation. Their tree-based semantic parsing model was trained on pairs of natural language and code mined from Stack Overflow and API documentation, and evaluation found that developer perception of the tool was positive.

In summary, source code search has been gaining traction in academia as researchers realise its ability to improve development productivity, specifically in cases involving the reuse of existing functionality over  the creation of new functionality. There is room to improve \tname{}'s code search, for example, we remain to evaluate how \tname{} can benefit from different data sources and semantic search.

\subsection{Alternative Library Selection}

El-Hajj and Nadi proposed LibComp~\cite{10.1145/3368089.3417922}, an IntelliJ plugin that assists the developer in selecting alternative libraries based on quality metrics associated with the libraries. LibComp does not perform code search. It requires humans to provide the association of libraries to its knowledge base. We are eager to evaluate the impact of using different metrics to sort the libraries reported by \tname. Also, the possibility of enabling \tname\ to search for alternatives (\eg{}, alternatives of a library function) seems promising. 

\hl{Additionally, there has been work on what factors influence the selection of packages}~\cite{larios2020selecting}.\hl{ Package features identified include documentation, performance, test coverage, security, maintenance, updates and size of community. Incorporating additional measures of package quality in \tname{} may improve the tool's ability to connect developers with suitable packages.}

\subsection{Code Executability}

Prior work identified different sources of problems when attempting to reproduce code available on the Web (e.g., tutorials~\cite{10.1145/3368089.3409706}) and proposed tools to improve code executability. For example, Mirhosseini and Parnin~\cite{10.1145/3368089.3409706} proposed a system to annotate documents and encouraged the use of notebooks (e.g., Jupyter Notebooks~\cite{jupyter}) to improve executability of tutorials. Similarly, Melo \etal{}~\cite{MeloETAL_TSE2019} proposed Frisk, a system that enables QA forum users to describe problems (e.g., configuration problems) and share these problems with the community (e.g., Stack Overflow users), who could help fix the problems. Frisk uses Docker to enable developers consistently reproduce problems and collaboratively propose fixes. However, unlike NCQ, Frisk does not try to automatically improve executability.

\section{Threats to Validity}
\label{sec:threats}

Similar to other empirical studies, there are threats which may affect the validity of our evaluation of \tname{}.

Threats to the \textit{construct validity} correspond to the appropriateness of the evaluation metrics. We evaluated \tname{} in terms of participant performance and participant perceptions. Our evaluation approaches, such as the Technology Acceptance Model, have been used before in similar studies (e.g.,~\cite{steinmacher2016overcoming}), and our evaluation methodology reflects our goals behind developing \tname{}.

Threats to the \textit{internal validity} compromise our confidence in establishing a relationship between the independent and dependent variables. Participants were not informed exactly how versions of \tname{} (\tnamei{}, \baseline{} and \tnameii{}) were implemented to ensure that their behaviour and responses would not be biased towards any of the tools. In both evaluations, participants worked on different tasks using both the tool and a baseline, which may have affected their experience and responses to our questions. The demonstration of how to use the tools at the beginning of each session may have influenced how participants used them; we mitigated this by having all participants watch the same demo video before their sessions. Because the study was conducted remotely, we cannot guarantee that participants did not search for solutions outside of the study setup. Participants were asked to use the web to find packages in task~1, however, were free to choose what websites to use; in this case, the choice of search engine, and thus knowledge of what options exist, may have influenced results. The design of our survey and participant screening was intended to minimise the number of non-programmer responses in our data, however, these measures may have had additional impacts on the participant pool.

Threats to \textit{external validity} correspond to the ability to generalise our results. We cannot claim generalisability beyond Node.js or the particular implementations of \tname{} and \baseline{} used in our study. Recruiting more or different developers to participate in the study and asking them to work on different tasks may have led to different results. \hl{We cannot conclude that improvements in performance between baseline and tool were due to any specific techniques or the automation of the process itself.}

\section{Conclusion and Future Work}

The results of our evaluation indicate that Node.js developers have distinct expectations for a code reuse tool, stemming from Node.js's package-based ecosystem. In general, we found that performance of participants in solving tasks did not reduce when using \tnameii{} compared to a more familiar baseline of VSCode and web access, despite the use of an unfamiliar tool with new commands and controls. We found that the perceptions participants had about \tnameii{} were generally positive, with most participants finding \tnameii{} helpful in accomplishing their assigned tasks. Participants' responses to questions about usefulness, easy of use and likely feature use using the TAM (Technology Acceptance Model) were also mostly positive. We also observed that \tnameii{}'s code correction features successfully improved the quality of snippets that participants interacted with.

However, our results also indicate that there is more that can still be done to aid developers in streamlining the code reuse process. \hl{Better measures of package quality could be investigated other than popularity and runnability.} Future work may also involve implementing NCQ in a different context, for example, many participants mentioned they would like an extension to their existing editor over a new tool. To enable package and code snippet search in an editor, we could implement NCQ as a plug-in for VSCode. 

As the Node.js ecosystem is always growing, the ability to update the tool's database would be useful (we use an offline database for efficiency). Currently, to update the database, the mining process must be re-run, but it may be possible to use the NPM registry API to keep the database up-to-date. As existing code search tools include online search, we did not investigate this aspect in this paper.

Future work could also expand on code snippet sources; the current version of the tool only uses examples from README, however, we found that just 38.6\% of NPM packages had code snippets in their READMEs. In these cases, mining API usage examples from NPM package source code may be useful.

We observed that when participants tried multiple packages, such as in the examples in Section~\ref{sec:example}, they were looking for similar packages; the package search thus could incorporate a `similar' package suggestion system, using package data to determine package `similarity'.

\section*{Acknowledgement}
This research was partially funded by INES 2.0, FACEPE grants PRONEX APQ 0388-1.03/14 and APQ-0399-1.03/17, CAPES grant 88887.136410/2017-00,  CNPq grant 465614/2014-0, and by ARC grants DP200102364 and DP210102670. Brittany’s research was supported by an Australian Government Research Training Program (RTP) Scholarship. The work of Christoph and Markus was supported by gifts from Facebook and Google.

\IEEEtriggeratref{22} 
\bibliographystyle{IEEEtran}
\bibliography{main.bib}

\end{document}

%% file: packages.tex
\usepackage[svgnames, table]{xcolor}
\usepackage[utf8]{inputenc}
\usepackage[english]{babel}
\usepackage{hyperref}
\hypersetup{
    colorlinks=true,
    linkcolor=blue,
    anchorcolor=blue,
    filecolor=blue,      
    citecolor=blue,
    urlcolor=blue
}
\usepackage{fancyvrb}
\usepackage{subcaption}
\usepackage{colortbl}
\usepackage{multirow}
\usepackage{subcaption}
\usepackage{listings}
\usepackage[list-final-separator={ e }, range-phrase={ até }]{siunitx}
\usepackage{comment}
\usepackage{adjustbox}
\usepackage{booktabs}
\usepackage[inline]{enumitem}
\usepackage{ textcomp }
\usepackage{wrapfig}
\usepackage{changepage}
\usepackage{standalone}
\usepackage{tikz}
\usetikzlibrary{
    arrows.meta,
    chains,
    scopes,
    positioning,
    shapes.geometric
}
\pgfdeclarelayer{background}
\pgfdeclarelayer{foreground}
\pgfsetlayers{background,main,foreground}

\definecolor{baseColour}{RGB}{255, 128, 85}
\definecolor{ncqColour}{RGB}{118, 165, 175}
\definecolor{ncqColour2}{RGB}{162, 196, 201}

\tikzset{
    processDiagram/.style={
        startstop/.style={
            rectangle,
            draw,
            minimum width=3cm, 
            minimum height=0.8cm,
            join = by arrow,
            fill = black!0,
        },
        decision/.style = {
            diamond,
            aspect=1.7,
            draw,
            minimum width=2.5cm, 
            minimum height=1cm, 
            align=center,
            join = by arrow,
            fill = black!0,
        },
        process/.style = {
            rectangle, 
            rounded corners, 
            draw,
            minimum width=3cm,
            minimum height=0.8cm, 
            align=center,
            join = by arrow,
            fill = black!0,
        },
        process2/.style = {
            rectangle, 
            rounded corners, 
            draw,
            minimum width=3cm,
            minimum height=0.8cm, 
            align=center,
            join = by arrow,
            fill = ncqColour2,
        },
        line/.style = {
            -
        },
        invisible/.style = {
            join = by line,
            minimum width=0cm, 
            minimum height=0cm,
            inner sep=0pt,
        },
        invisititle/.style ={
            font = \bfseries,
            join = by arrow,
            outer sep = 5pt,
        },
        title/.style = {
            font = \bfseries,
            fill = black!20,
            join = by arrow,
            minimum width = 3cm,
            minimum height=0.8cm,
            draw,
        },
        output/.style ={
            join = by line,
            minimum width=0cm, 
            minimum height=0cm,
            inner sep=0pt,
            font = \itshape,
        },
        arrow/.style = {
            ->
        },
    }
}

\usepackage{color}
\newcommand{\hl}[1]{#1}
\definecolor{darkblue}{rgb}{0,0,0.5}
\definecolor{LightGray}{rgb}{0.75,0.75,0.75}
\definecolor{VeryLightGray}{rgb}{0.90,0.90,0.90}
\definecolor{ForestGreen}{rgb}{0.13,0.54,0.13}
\definecolor{lightgreen}{rgb}{0.8,1,0.8}
\definecolor{lightyellow}{rgb}{1,1,0.8}
\definecolor{lightred}{rgb}{1,0.8,0.8}

\lstdefinelanguage{diff}{
    morecomment=[f][\lstbg{lightgreen}]+,
    morecomment=[f][\lstbg{lightred}]-,
    morecomment=[f][\lstbg{lightyellow}]\\,
}
\newcommand{\lstbg}[3][0pt]{{\fboxsep#1\colorbox{#2}{\strut #3}}}
\usepackage{cite} 
\usepackage[noend]{algpseudocode}
\usepackage{algorithm}
\usepackage{tikz}
\usetikzlibrary{matrix,chains,positioning,decorations.pathreplacing,arrows}
\usepackage[skins]{tcolorbox}
\usepackage{tabularx}
\usepackage{nicefrac}

%% file: macros.tex
\newcommand{\numparticipants}{10}

\newcommand{\CodeIn}[1]{\begin{small}\texttt{#1}\end{small}}

\newcommand{\Ignore}[1]{}
\newcommand{\eg}{e.g.}
\newcommand{\ie}{i.e.}

\newcommand{\etal}{et al.}

\newcommand{\tname}{NCQ}
\newcommand{\tnamei}{NCQ$_{Retrieval}$}
\newcommand{\tnameii}{NCQ}
\newcommand{\baseline}{NCQ$_{Basic}$}

\newcommand{\javascript}{JavaScript}
\newcommand{\stackoverflow}{Stack Overflow}

\newcommand{\Space}[1]{}
\newcommand{\Contrib}[1]{$\star$#1}

%% file: figures/ncqrfeatures.tex
\begin{tikzpicture}[processDiagram, node distance=4mm and 5cm]
{ [start chain=trunk going below]
    \node [startstop, on chain] (n0) {Node.js REPL};
    {[start branch = e going right]
        \node [title, on chain, right = 1] (e0) {(1) Environment Setup};
    }
    {[start branch = e going right]
        \node[invisible, on chain, right = 0.5] {};
        \node[invisible, on chain, below = 1] {};
        \node [title, on chain, right = 1.1] (e1) {(2) Code Execution};
    }
    \node [decision, on chain, below = 1] (commands) {Commands};
    {[start branch = p going below]
        \node[invisible, on chain = going left, xshift = 20] {};
        \node[invisititle, on chain,  below=1.4] (p0) {(4) Package Search};
        \node[process, on chain] (p1) {.packages $<$query$>$};
        \node[output, on chain] {.packages connect to sql};
        \node[process, on chain] (p2) {Process into keywords};
        \node[output, on chain] {[``connect",``sql"]};
        \node[process, on chain] (p3) {Search database by keywords};
        \node[output, on chain] {[set of retrieved packages]};
        \node[process, on chain] {Sort by stars};
        \node[output, on chain] {[sorted set]};
        \node[process, on chain] (p9) {Print table};
    }
    {[start branch = s going below]
        \node[invisititle, on chain,  below=0.6] (s0) {(5) Code Snippet Search};
        \node[process, on chain] (s1) {.samples $<$package$>$};
        \node[output, on chain] {.packages mysql};
        \node[process, on chain] (s2) {Search code snippet \\ database};
        \node[output, on chain] (so3) {[set of retrieved code snippets]};
        \node[process, on chain] {Sort by order in README};
        \node[output, on chain] {[sorted set]};
        \node[process, on chain] (s7) {Autofill input \\ with code snippet};
    }
    {[start branch = packages going below]
        \node[title, on chain = going right, xshift=-90] (e2) {(3) Editor Mode};
    }
}
\begin{pgfonlayer}{background}
    \path (n0.west |- n0.north)+(-0.2, 0.2) node (e) {};
    \path (e0.east |- e2.south)+(0.2, -0.55) node (f) {};
    \path[fill=baseColour, draw](e) rectangle (f); 
    \node[below of = e2, yshift=-10, xshift = 20]{\textbf{NCQ$_{Basic}$}};

    \path (p3.west |- p0.north)+(-0.2,0) node (a) {};
    \path (p3.east |- p9.south)+(0.2, -0.2) node (b) {};
    \path[fill=black!20, draw](a) rectangle (b); 
    
     \path (so3.west |- s0.north)+(-0.2,0) node (c) {};
    \path (so3.east |- s7.south)+(0.2,-0.5) node (d) {};
    \path[fill=black!20, draw](c) rectangle (d); 
\end{pgfonlayer}

\end{tikzpicture}

%% file: figures/ncqfeatures.tex
\begin{tikzpicture}[processDiagram, node distance=4mm and 5cm]
{ [start chain=trunk going below]
    \node [startstop, on chain] (n0) {Node.js REPL};
    {[start branch = e going right]
        \node [title, on chain, right = 1] (e0) {(1) Environment Setup};
    }
    {[start branch = e going right]
        \node[invisible, on chain, right = 0.5] {};
        \node[invisible, on chain, below = 1] {};
        \node [title, on chain, right = 1.1] (e1) {(2) Code Execution};
    }
    \node [decision, on chain, below = 1] (commands) {Commands};
    {[start branch = p going below]
        \node[invisible, on chain = going left] {};
        \node[invisititle, on chain,  below=1.4] (p0) {(4) Package Search};
        \node[process, on chain] (p1) {.packages $<$query$>$};
        \node[output, on chain] {.packages connect to sql};
        \node[process, on chain] (p2) {Process into keywords};
        \node[output, on chain] {[``connect",``sql"]};
        \node[process, on chain] (p3) {Search database by keywords};
        \node[output, on chain] {[set of retrieved packages]};
        \node[invisititle, on chain] (i0) {(N1) Improved Package Search};
        \node[process2, on chain] {Sort by ability to build};
        \node[output, on chain] {[sorted set]};
        \node[process2, on chain] {Sort by has stars};
        \node[output, on chain] {[sorted set]};
        \node[process2, on chain] (p9) {\textbf{(N2) Interactive Package List}};
    }
    {[start branch = s going below]
        \node[invisititle, on chain,  below=0.6] (s0) {(5) Code Snippet Search};
        \node[process, on chain] (s1) {.samples $<$package$>$};
        \node[output, on chain] {.packages mysql};
        \node[process, on chain] {Search code snippet \\ database};
        \node[output, on chain] (so3) {[set of retrieved code snippets]};
        \node[invisititle, on chain] (c0) {(N3) Code Correction};
        \node[process2, on chain] {Run ESLint};
        \node[output, on chain] {errors, snippet};
        \node [decision, on chain] (errors) {Parsing  \\ Errors?};
        {[start branch = p going right]
            \node[output, on chain, right=0.2] {yes};
            \node[process2, on chain, right = 0.4] {Line deletion};
            \node[process2, on chain = going below](final) {Get final error count};
        }
        \node[output, on chain] {no};
        \node[process2, on chain] (c3) {Sort by errors};
        \node[output, on chain] {[sorted set]};
        \node[process, on chain] (s7) {Autofill input \\ with code snippet};
    }
    {[start branch = packages going below]
        \node[title, on chain = going right, xshift=-90] (e2) {(3) Editor Mode};
    }
}

\draw[arrow](final) to [bend left=20] (c3);
\begin{pgfonlayer}{background}
    \path (p3.west |- p0.north)+(-0.8,0) node (a) {};
    \path (p3.east |- p9.south)+(+0.8,-0.4) node (b) {};
    \path[fill=black!20, draw](a) rectangle (b); 
    
    \path (i0.west |- i0.north)+(0.15,0) node (a) {};
    \path (i0.east |- p9.south)+(-0.15,-0.2) node (b) {};
    \path[fill=ncqColour, draw](a) rectangle (b); 
    
     \path (so3.west |- s0.north)+(-0.6,0) node (c) {};
    \path (so3.east |- s7.south)+(0.6,-0.2) node (d) {};
    \path[fill=black!20, draw](c) rectangle (d); 
    
    \path (c0.west |- c0.north)+(-0.3,0) node (a) {};
    \path (c0.east |- c3.south)+(4.6,-0.2) node (b) {};
    \path[fill=ncqColour, draw](a) rectangle (b); 
    
    \path (n0.west |- n0.north)+(-0.2, 0.2) node (e) {};
    \path (e0.east |- e2.south)+(0.2, -0.55) node (f) {};
    \path[fill=baseColour, draw](e) rectangle (f); 
    \node[below of = e2, yshift=-10, xshift = 20]{\textbf{NCQ$_{Basic}$}};
\end{pgfonlayer}

\end{tikzpicture}

%% file: main.bbl
\begin{thebibliography}{10}
\providecommand{\url}[1]{#1}
\csname url@samestyle\endcsname
\providecommand{\newblock}{\relax}
\providecommand{\bibinfo}[2]{#2}
\providecommand{\BIBentrySTDinterwordspacing}{\spaceskip=0pt\relax}
\providecommand{\BIBentryALTinterwordstretchfactor}{4}
\providecommand{\BIBentryALTinterwordspacing}{\spaceskip=\fontdimen2\font plus
\BIBentryALTinterwordstretchfactor\fontdimen3\font minus
  \fontdimen4\font\relax}
\providecommand{\BIBforeignlanguage}[2]{{%
\expandafter\ifx\csname l@#1\endcsname\relax
\typeout{** WARNING: IEEEtran.bst: No hyphenation pattern has been}%
\typeout{** loaded for the language `#1'. Using the pattern for}%
\typeout{** the default language instead.}%
\else
\language=\csname l@#1\endcsname
\fi
#2}}
\providecommand{\BIBdecl}{\relax}
\BIBdecl

\bibitem{nodejs_stats}
\BIBentryALTinterwordspacing
{W3Techs}, ``{Usage Statistics of Node.js},'' 2020. [Online]. Available:
  \url{https://w3techs.com/technologies/details/ws-nodejs}
\BIBentrySTDinterwordspacing

\bibitem{snyk_io_blog}
\BIBentryALTinterwordspacing
L.~Tal and S.~Maple, ``{npm passes the 1 millionth package milestone! What can
  we learn?}'' 2020. [Online]. Available:
  \url{https://snyk.io/blog/npm-passes-the-1-millionth-package-milestone-what-can-we-learn/}
\BIBentrySTDinterwordspacing

\bibitem{kula2017impact}
R.~G. Kula, A.~Ouni, D.~M. German, and K.~Inoue, ``On the impact of
  micro-packages: An empirical study of the npm javascript ecosystem,''
  \emph{arXiv:1709.04638}, 2017.

\bibitem{sandewall1978programming}
E.~Sandewall, ``Programming in an interactive environment:
  the``lisp''experience,'' \emph{ACM Computing Surveys (CSUR)}, vol.~10, no.~1,
  pp. 35--71, 1978.

\bibitem{Campbell2017NLP2CodeCS}
B.~A. Campbell and C.~Treude, ``{NLP2Code}: Code snippet content assist via
  natural language tasks,'' in \emph{ICSME}, 2017, pp. 628--632.

\bibitem{ponzanelli2014prompter}
L.~Ponzanelli, G.~Bavota, M.~Di~Penta, R.~Oliveto, and M.~Lanza, ``Prompter: A
  self-confident recommender system,'' in \emph{ICSME}, 2014, pp. 577--580.

\bibitem{zhang2016bing}
H.~Zhang, A.~Jain, G.~Khandelwal, C.~Kaushik, S.~Ge, and W.~Hu, ``Bing
  developer assistant: improving developer productivity by recommending sample
  code,'' in \emph{FSE}, 2016, pp. 956--961.

\bibitem{brandt2010example}
J.~Brandt, M.~Dontcheva, M.~Weskamp, and S.~R. Klemmer, ``Example-centric
  programming: integrating web search into the development environment,'' in
  \emph{Conference on Human Factors in Computing Systems (CHI)}, 2010, pp.
  513--522.

\bibitem{ponzanelli2013seahawk}
L.~Ponzanelli, A.~Bacchelli, and M.~Lanza, ``Seahawk: Stack overflow in the
  ide,'' in \emph{ICSE}.\hskip 1em plus 0.5em minus 0.4em\relax IEEE, 2013, pp.
  1295--1298.

\bibitem{10.1145/3368089.3417922}
R.~El-Hajj and S.~Nadi, ``{LibComp: An IntelliJ Plugin for Comparing Java
  Libraries},'' in \emph{ACM Joint European Software Engineering Conference and
  Symposium on the Foundations of Software Engineering (ESEC/FSE)}, 2020, p.
  1591–1595.

\bibitem{NLP2TestableCode}
B.~Reid, C.~Treude, and M.~Wagner, ``Optimising the fit of stack overflow code
  snippets into existing code,'' in \emph{GECCO}, 2020, p. 1945–1953.

\bibitem{doyoucode}
A.~Danilova, A.~Naiakshina, S.~Horstmann, and M.~Smith, ``Do you really code?
  designing and evaluating screening questions for online surveys with
  programmers,'' in \emph{ICSE}, 2021, pp. 537--548.

\bibitem{prolificUserStudy}
B.~Reid, M.~Wagner, M.~d'Amorim, and C.~Treude, ``Software engineering user
  study recruitment on prolific: An experience report,'' 2022.

\bibitem{framework}
\BIBentryALTinterwordspacing
A.~Rainer and C.~Wohlin, ``Recruiting credible participants for field studies
  in software engineering research,'' 2021. [Online]. Available:
  \url{https://arxiv.org/abs/2112.14186}
\BIBentrySTDinterwordspacing

\bibitem{TAMdavis1989perceived}
F.~D. Davis, ``Perceived usefulness, perceived ease of use, and user acceptance
  of information technology,'' \emph{Management Information Systems (MIS)
  Quarterly}, pp. 319--340, 1989.

\bibitem{TAMvenkatesh2000theoretical}
V.~Venkatesh and F.~D. Davis, ``A theoretical extension of the technology
  acceptance model: Four longitudinal field studies,'' \emph{Management
  Science}, vol.~46, no.~2, pp. 186--204, 2000.

\bibitem{Proksch2015}
S.~Proksch, V.~Bauer, and G.~C. Murphy, \emph{How to Build a Recommendation
  System for Software Engineering}.\hskip 1em plus 0.5em minus 0.4em\relax
  Springer, 2015, pp. 1--42.

\bibitem{chinthanet2021makes}
B.~Chinthanet, B.~Reid, C.~Treude, M.~Wagner, R.~G. Kula, T.~Ishio, and
  K.~Matsumoto, ``What makes a good node.js package? investigating users,
  contributors, and runnability,'' 2021.

\bibitem{campos2019mining}
U.~F. Campos, G.~Smethurst, J.~P. Moraes, R.~Bonif{\'a}cio, and G.~Pinto,
  ``Mining rule violations in javascript code snippets,'' in \emph{MSR}.\hskip
  1em plus 0.5em minus 0.4em\relax IEEE, 2019, pp. 195--199.

\bibitem{abdalkareem2017developers}
R.~Abdalkareem, O.~Nourry, S.~Wehaibi, S.~Mujahid, and E.~Shihab, ``Why do
  developers use trivial packages? an empirical case study on npm,'' in
  \emph{ACM Joint European Software Engineering Conference and Symposium on the
  Foundations of Software Engineering (ESEC/FSE)}, 2017, p. 385–395.

\bibitem{basili1996reuse}
V.~R. Basili, L.~C. Briand, and W.~L. Melo, ``How reuse influences productivity
  in object-oriented systems,'' \emph{Communications of the ACM (CACM)},
  vol.~39, no.~10, pp. 104--116, 1996.

\bibitem{mockus2007large}
A.~Mockus, ``Large-scale code reuse in open source software,'' in
  \emph{International Workshop on Emerging Trends in FLOSS Research and
  Development}.\hskip 1em plus 0.5em minus 0.4em\relax IEEE, 2007, pp. 7--7.

\bibitem{lim1994effects}
W.~C. Lim, ``Effects of reuse on quality, productivity, and economics,''
  \emph{IEEE Software}, vol.~11, no.~5, pp. 23--30, 1994.

\bibitem{mohagheghi2004empirical}
P.~Mohagheghi, R.~Conradi, O.~M. Killi, and H.~Schwarz, ``An empirical study of
  software reuse vs. defect-density and stability,'' in \emph{ICSE}, 2004, pp.
  282--291.

\bibitem{sojer2010code}
M.~Sojer and J.~Henkel, ``Code reuse in open source software development:
  Quantitative evidence, drivers, and impediments,'' \emph{Journal of the
  Association for Information Systems (JAIS)}, vol.~11, no.~12, pp. 868--901,
  2010.

\bibitem{li2016relationship}
X.~Li, Z.~Wang, Q.~Wang, S.~Yan, T.~Xie, and H.~Mei, ``Relationship-aware code
  search for javascript frameworks,'' in \emph{FSE}, 2016, pp. 690--701.

\bibitem{mcmillan2010recommending}
C.~McMillan, D.~Poshyvanyk, and M.~Grechanik, ``Recommending source code
  examples via api call usages and documentation,'' in \emph{International
  Workshop on Recommendation Systems for Software Engineering}, 2010, pp.
  21--25.

\bibitem{moreno2015can}
L.~Moreno, G.~Bavota, M.~Di~Penta, R.~Oliveto, and A.~Marcus, ``How can i use
  this method?'' in \emph{ICSE}, vol.~1.\hskip 1em plus 0.5em minus 0.4em\relax
  IEEE, 2015, pp. 880--890.

\bibitem{SAIED2018164}
M.~A. Saied, A.~Ouni, H.~Sahraoui, R.~G. Kula, K.~Inoue, and D.~Lo, ``Improving
  reusability of software libraries through usage pattern mining,''
  \emph{Journal of Systems and Software}, vol. 145, pp. 164--179, 2018.

\bibitem{gu2018deep}
X.~Gu, H.~Zhang, and S.~Kim, ``Deep code search,'' in \emph{ICSE}.\hskip 1em
  plus 0.5em minus 0.4em\relax IEEE, 2018, pp. 933--944.

\bibitem{inIDECodeGen}
\BIBentryALTinterwordspacing
F.~F. Xu, B.~Vasilescu, and G.~Neubig, ``In-ide code generation from natural
  language: Promise and challenges,'' \emph{TOSEM}, vol.~31, no.~2, mar 2022.
  [Online]. Available: \url{https://doi.org/10.1145/3487569}
\BIBentrySTDinterwordspacing

\bibitem{DBLP:conf/icml/KanadeMBS20}
A.~Kanade, P.~Maniatis, G.~Balakrishnan, and K.~Shi, ``Learning and evaluating
  contextual embedding of source code,'' in \emph{International Conference on
  Machine Learning (ICML)}, vol. 119, 2020, pp. 5110--5121.

\bibitem{larios2020selecting}
E.~Larios~Vargas, M.~Aniche, C.~Treude, M.~Bruntink, and G.~Gousios,
  ``Selecting third-party libraries: The practitioners’ perspective,'' in
  \emph{Proceedings of the 28th ACM joint meeting on european software
  engineering conference and symposium on the foundations of software
  engineering}, 2020, pp. 245--256.

\bibitem{10.1145/3368089.3409706}
S.~Mirhosseini and C.~Parnin, ``Docable: Evaluating the executability of
  software tutorials,'' in \emph{ACM Joint European Software Engineering
  Conference and Symposium on the Foundations of Software Engineering
  (ESEC/FSE)}, 2020, p. 375–385.

\bibitem{jupyter}
\BIBentryALTinterwordspacing
``Jupyter,'' 2020. [Online]. Available: \url{https://jupyter.org/}
\BIBentrySTDinterwordspacing

\bibitem{MeloETAL_TSE2019}
L.~{Melo}, I.~S. {Wiese}, and M.~{d'Amorim}, ``{Using Docker to Assist Q A
  Forum Users},'' \emph{TSE}, 2019.

\bibitem{steinmacher2016overcoming}
I.~Steinmacher, T.~U. Conte, C.~Treude, and M.~A. Gerosa, ``Overcoming open
  source project entry barriers with a portal for newcomers,'' in \emph{ICSE},
  2016, pp. 273--284.

\end{thebibliography}
